\documentclass[reprint,superscriptaddress,amsmath,amssymb,amsthm,aps,prx,nofootinbib]{revtex4-2}

\usepackage{graphicx}
\usepackage{bm}
\usepackage{enumitem}
\usepackage[version=4]{mhchem}
\usepackage{inconsolata}
\usepackage{tikz}
\usepackage{setspace}
\usepackage{titletoc}
\usepackage[ruled,vlined,longend,algo2e]{algorithm2e}

\usepackage[table,dvipsnames]{xcolor}
\definecolor{linkcolor}{HTML}{b91c1c}
\definecolor{citecolor}{HTML}{64748b}
\definecolor{urlcolor}{HTML}{2563eb}
\definecolor{liflow}{HTML}{ede9fe}

\usepackage{booktabs}
\usepackage{tabularray}
\UseTblrLibrary{booktabs}
\usepackage{makecell}
\usepackage{multirow}

\usepackage{hyperref}
\hypersetup{colorlinks,linkcolor=linkcolor,citecolor=citecolor,urlcolor=urlcolor}
\bibpunct{\color{citecolor}[}{\color{citecolor}]}{,}{n}{}{;}  % citation bracket
\usepackage[capitalise]{cleveref}

\usepackage{amsthm}

\renewcommand{\figurename}{Fig.}
\renewcommand{\tablename}{Table}
\renewcommand{\thetable}{\arabic{table}}

\crefname{extendedfigure}{Extended Data Fig.}{Extended Data Figs.}
\crefname{extendedtable}{Extended Data Table}{Extended Data Tables}

\crefname{suppfigure}{Fig.}{Figs.}
\crefname{supptable}{Table}{Tables}
\crefname{suppsection}{Supplementary Note Section}{Supplementary Note Sections}

\begin{document}

%\title{Universal Markov-Chain-Based Algorithms for Decomposing Onsager Coefficients to Quantify Transport Mechanisms in Ion Conducting Media}
%\title{Universal framework for decomposing discrete interpretable ionic transport mechanisms}
\title{Universal Framework for Decomposing Ionic Transport into Interpretable Mechanisms}
%\title{Universal discrete stochastic framework for decomposing ionic transport mechanisms}
\author{KyuJung Jun}
\thanks{These authors contributed equally}
\author{Pablo A. Leon}
\thanks{These authors contributed equally}
\author{Jur\u{g}is Ru\v{z}a}
\author{Juno Nam}
\author{Rafael G\'omez-Bombarelli}
\email{rafagb@mit.edu}
\affiliation{Department of Materials Science and Engineering, Massachusetts Institute of Technology, Cambridge, MA 02139, USA}

\date{\today}

\begin{abstract}
Understanding mechanisms of ion transport in bulk materials is central to designing next-generation ion conductors for energy storage devices, yet studies employing all-atom molecular dynamics (MD) have largely been limited to reporting overall transport coefficients without a quantitative, spatiotemporally resolved breakdown of \emph{how} charge is carried. We present a computational framework that analyzes MD trajectories to quantitatively interpret macroscopic transport by decomposing it into additive contributions from physically motivated events. They are defined either through heuristically identified microscopic transitions, capturing events such as single-ion hops, multi-ion hops, and vehicular motion, or through transitions between chemically interpretable coordination macrostates. The construction guarantees that attributed contributions sum exactly to the Onsager transport coefficients estimated via the Green–Kubo/Einstein formalism, while scanning the sampling window exposes characteristic temporal scales at which distinct transport mechanisms emerge and dominate. Applied across three prototypical electrolytes—inorganic crystals, liquids, and polymers—the framework quantitatively resolves long-standing debates (e.g., the role of concerted motion and exchange), identifies dominant mechanisms and rate-limiting steps, quantifies their frequencies and effectiveness, and extracts activation energies for distinct transport modes, thereby distilling design rules for fast conduction. This general and reproducible analysis tool turns MD trajectories into quantitative mechanism maps, enabling the ion-conductor community to adjudicate mechanistic hypotheses and accelerate discovery.

\end{abstract}

\maketitle

%%%%%%%%%%%%%%%%%%%%%%%%%%%%%%%%%%%%%%%%%%%%%%%%%%%%%%%%%%%%%%%%%%%%%%%%%%%%%%%%
% Introduction
%%%%%%%%%%%%%%%%%%%%%%%%%%%%%%%%%%%%%%%%%%%%%%%%%%%%%%%%%%%%%%%%%%%%%%%%%%%%%%%%

\section{Introduction}
\label{sec:introduction}
Ion transport through bulk materials is central to the design of next-generation ion conductors for energy storage devices\citep{MengXuScience2022}. In batteries, the electrolyte ionic conductivity, cation transference number, and activation energy barriers directly govern internal resistance, operating temperature\citep{hubble2022liquid-48b}, rate capability\citep{jiang2016elucidating-024}, and limiting currents\citep{shah2019comparing-741, newman2021electrochemical-304, lee2024experimental-39a}. To meet growing performance and safety targets for next generation batteries, the community is exploring a wide landscape of ionic conductors: tuned liquid electrolytes that vary salt concentration (including localized high concentration electrolytes and solvate ionic liquids)\citep{hubble2022liquid-48b, xu2004nonaqueous-a2a}, solvent/anion chemistries\citep{lim2023electrolyte-e3a}, and cosolvents or diluents to modulate solvation structure and viscosity\citep{wang2022liquid-1ad}; inorganic crystalline solid electrolytes (oxides, sulfides, halides)\citep{zhao2020designing-72d, JanekZeierNatureEnergy2016}; and \emph{soft-matter} systems\citep{zhou2019polymer-b39} such as polymer electrolytes, gels\citep{liu2025bioinspired-1d6}, charged or zwitterionic polymers\citep{jones2022design-721}, and supramolecular electrolytes\citep{bae2023closed-loop-128}, and their composites\citep{zhang2023review-75f}. While design priorities vary from (electro)chemical stability to safety and mechanical robustness, optimizing ion-transport properties such as ionic conductivity and cation transference number remains essential across all material classes. Going beyond trial and error, however, requires a rigorous \emph{mechanistic} understanding of transport, and thus discussions inevitably turn to \emph{how} ions move: which microscopic motions actually carry ions, on what spatiotemporal scales, and how chemical and structural design can promote accelerated transport mechanisms\citep{jun2024diffusion-7a6, ChooBalsaraProgressinPolymerScience2020}.

Experimentally, macroscopic transport properties are routinely measured by electrochemical impedance spectroscopy (EIS)\citep{lazanas2023electrochemical-ca5, LaiHaileJournaloftheAmericanCeramicSociety2005}, pulsed-field gradient (PFG) and electrophoretic nuclear magnetic resonance (NMR) (including \textsuperscript{7}Li relaxometry)\citep{KuhnHeitjansSolidStateNuclearMagneticResonance2012, KuhnLotschEnergyEnvironmentalScience2013, hickson2022complete-2a7}, and quasi-elastic neutron scattering (QENS)\citep{KlenkLaiSolidStateIonics2017}, among other techniques. These measurements, however, cannot directly reveal \emph{how} charged particles move at \r{A}ngstr\"{o}m and picosecond scales \emph{simultaneously}, since experimental techniques typically achieve high resolution in either space or time, but rarely both at once\citep{GaoBoChemicalReviews2020}. All-atom molecular dynamics (MD) simulations inherently offer both high spatial and temporal resolution, resolving atomic-scale structure and picosecond-scale dynamics. Yet, in practice, MD has largely been limited to reporting \emph{overall} transport coefficients or highlighting a few representative snapshots of particular transport events\citep{hou2021solvation-ec1, brooks2018atomistic-6b2, JalemKanamuraChemistryofMaterials2013}. A long-standing and notoriously challenging computational goal has been to develop a quantitative, spatiotemporally resolved mechanistic attribution of transport that provides a principled way to determine which processes dominate at which timescales, how their contributions combine to yield the observed macroscopic transport coefficients, and how effective each mode is at carrying ions.

Existing analyses address parts of this question but remain fragmented, system-specific, and difficult to generalize, making mechanistic interpretation of simulations a bottleneck for closed-loop computation-driven design. In inorganic conductors, minimum-energy pathway methods such as nudged elastic band (NEB)\citep{henkelman2002theoretical-d66} identify candidate hops and activation barriers, but they do not quantify how frequently each pathway occurs at finite temperature or how much each contributes to macroscopic transport. In liquids and amorphous solids, the number of possible transitions becomes combinatorially large, rendering such approaches intractable. In crystalline materials or amorphous phases such as polymer electrolytes, microscopic or kinetic models using Monte Carlo schemes parameterized from MD can reproduce self-diffusion, but they require tailored model construction for each system, rely on simplifying assumptions, and do not guarantee that the sum of contributions from each mechanism equals the exact transport coefficient obtained from MD\citep{maitra2007cation-fa0, borodin2006mechanism-3ef, diddens2010understanding-ba4}. In liquid electrolytes, solvation-state classifications such as solvent-separated ion pair (SSIP), contact ion pair (CIP), and aggregates (AGG) describe thermodynamic populations, while vehicular and exchange processes describe kinetic transitions, but quantitatively connecting them to macroscopic transport remains challenging\citep{GaoBoChemicalReviews2020}.

Here we introduce \emph{OnsagerDecomposer}, a general, discrete stochastic framework that takes equilibrium MD trajectories sampled at multiple frequencies and yields timescale-resolved, additive attributions of macroscopic transport. The algorithm discretizes each ion’s motion into sampling windows of length $\Delta t$, describes the ion’s state at each snapshot by its coordination environment, and detects changes every $\Delta t$. These changes are mapped to a predefined, complete and nonoverlapping set of transitions to model ion transport as a discrete, event-labeled stochastic process. A simple example is classifying each transition as either a structural or vehicular transport event. At a coarser level, transitions can also occur between chemically defined \emph{macrostates} describing solvation environments, such as solvent-separated ion pairs (SSIP), contact ion pairs (CIP), or aggregates (AGG). By combining these time sequences of events and their corresponding displacements, the algorithm computes mechanism-specific contributions to the Onsager transport coefficients, which sum exactly to the conventional undecomposed values. By scanning across a range of $\Delta t$ values, we uncover the characteristic timescales over which distinct transport mechanisms emerge, dominate, and coalesce, providing a quantitative view of ion transport across temporal regimes.

We demonstrate the framework across three prototypical lithium-ion conductors—an inorganic crystalline electrolyte, a polymer electrolyte, and a liquid electrolyte—each representing a distinct class of ion-transport physics. In each case, \textit{OnsagerDecomposer} quantitatively decomposes ion motion into physically interpretable modes that correspond to long-standing mechanistic hypotheses: concerted versus single-ion hops in crystals, interchain versus intrachain (and vehicular) motion in polymers, and vehicular versus exchange-driven transport in liquids. By resolving how the \emph{frequency} and \emph{effectiveness} of these mechanisms evolve across timescales, the framework provides a rigorous, quantitative picture of ion transport in each system, revealing which microscopic processes dominate or limit ion transport across different spatiotemporal regimes.

Beyond these demonstrations, \textit{OnsagerDecomposer} provides a general and extensible framework for quantitatively linking atomic-scale dynamics to macroscopic transport. Its event- and macrostate-based definitions are both physically motivated and user-definable, allowing researchers to test mechanistic hypotheses, compare materials on equal footing, and uncover transferable design principles for fast ion conduction. By identifying which microscopic mechanisms most effectively promote conductivity and how their prevalence and efficacy evolve with chemistry, composition, and structure, the framework bridges atomistic dynamics and macroscopic performance. This unified, quantitative understanding paves the way for mechanism-aware, \textit{in-silico} design of next-generation ion conductors across liquids, polymers, and crystalline solids.

%%%%%%%%%%%%%%%%%%%%%%%%%%%%%%%%%%%%%%%%%%%%%%%%%%%%%%%%%%%%%%%%%%%%%%%%%%%%%%%%
% Results
%%%%%%%%%%%%%%%%%%%%%%%%%%%%%%%%%%%%%%%%%%%%%%%%%%%%%%%%%%%%%%%%%%%%%%%%%%%%%%%%

\section{Results}
\label{sec:results}

\subsection{Algorithmic description}
\label{sec:algorithmic_description}

\begin{figure*}[!ht]
\includegraphics[width=0.95\textwidth]{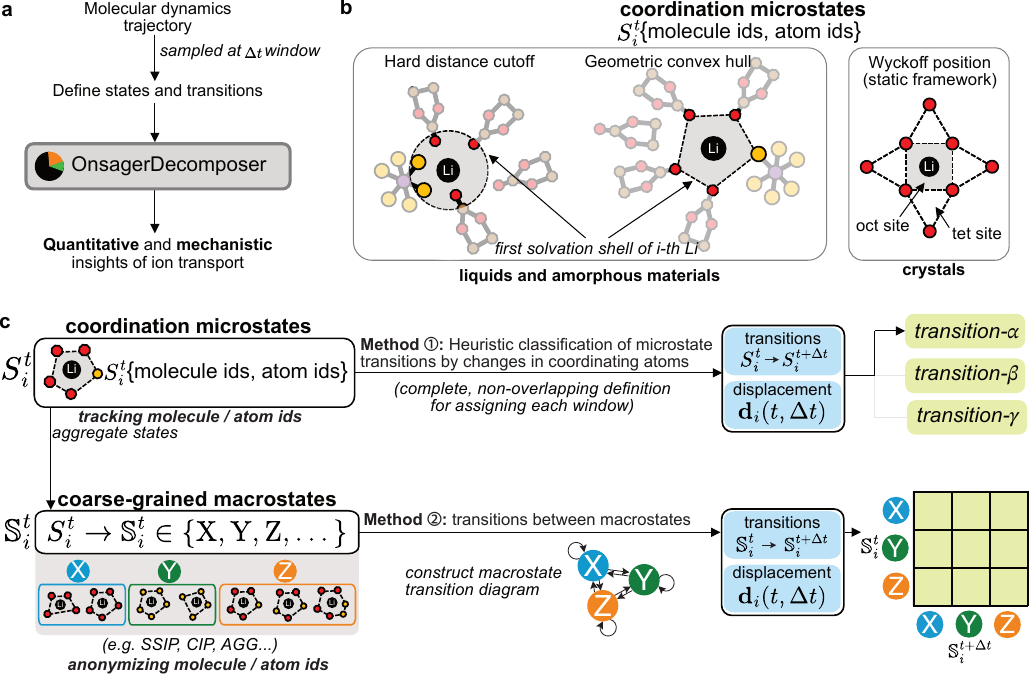}
\caption{
\textbf{Overview of \textit{OnsagerDecomposer}}
\textbf{a}, \textit{OnsagerDecomposer} processes equilibrium MD trajectories sampled at a window length $\Delta t$ to yield a mechanism-resolved \emph{decomposition} of Onsager transport coefficients.
\textbf{b}, Coordination microstates \(S_i^t\) encode the identities and indices of solvating species within the coordination shell of the ion of interest (Li) with index $i$ at time $t$, determined by hard distance cutoff or a geometric convex hull method for liquids and amorphous materials, and by mapping to unique Wyckoff positions in crystals. 
\textbf{c}, Mapping transitions. \emph{Method 1 (heuristic classification)} assigns each microstate transition \(S_i^t\!\to\!S_i^{t{+}\Delta t}\) and its corresponding displacement \(\mathbf{d}_i(t,\Delta t)\) to a complete, non-overlapping set of event classes \(\{m\}\) (e.g., vehicular vs exchange; intrachain vs interchain; single-ion vs concerted), shown schematically as \(\alpha,\beta,\gamma\). \emph{Method 2 (macrostate coarse-graining)} aggregates high-dimensional microstates to chemically interpretable macrostates, \(S_i^t\mapsto\mathbb{S}_i^t\) (e.g., SSIP, CIP, AGG; activated vs non-activated), labeled X, Y, Z. Each macrostate transition and its corresponding displacement \(\mathbf{d}_i(t,\Delta t)\) are then mapped into the macrostate transition diagram.
}\label{fig:overview}
\end{figure*}

\paragraph*{Objective.}
Given an equilibrium MD trajectory, our goal is to obtain a \emph{quantitative, timescale-resolved, additive decomposition} of macroscopic transport into (i) physically motivated microscopic \emph{events} that connect microstates and (ii) transitions between chemically interpretable \emph{macrostates} (Figure\ref{fig:overview}\textbf{a}). To understand the transport mechanism of ion type $A$ (e.g., Li), we analyze its motion as a function of the sampling window $\Delta t$, producing decomposed Onsager coefficients that each correspond to distinct transport modes, along with the \emph{effectiveness} of each mode, macrostate populations, and transition probabilities. The sampling frequency of the MD trajectory determines the accessible timescales, thereby revealing how different transport modes contribute across short- and long-time regimes. The terms \emph{events}, \emph{transitions}, and \emph{transport modes} are used interchangeably throughout this work.

\paragraph*{Defining microstates and their transitions.}
We analyze trajectories obtained by subsampling the MD trajectory at intervals of $\Delta t$. For the species of interest (ion type $A$), the displacement of ion $i$ over the window $[t, t + \Delta t]$ is defined as
\begin{equation}
\label{eq:disp}
\mathbf{d}_i(t,\Delta t) = \mathbf{r}_i(t{+}\Delta t) - \mathbf{r}_i(t),
\end{equation}
where $\mathbf{r}_i(t)$ is the position of ion $i$ at time $t$. A coordination microstate, $S_i^t$, encodes the identities (molecule and atom IDs) of species within the first coordination shell of ion $i$ at time $t$. We employ flexible methods to determine the coordination environment. As shown in Fig.\ref{fig:overview}\textbf{b}, shells are detected either by a hard distance cutoff at the first minimum of the radial distribution function, a parameter-free convex-hull criterion in liquids and polymers, or by Wyckoff-site assignment in crystals. We explicitly track microstates to distinguish between identical molecular species with different IDs, allowing us to identify when specific molecules enter or leave the coordination shell.

Based on heuristics appropriate to each class of materials, we define a set of \emph{events} or \emph{transitions} that each ion can undergo within a given time window. Representative cases are \emph{vehicular} motion in liquid electrolytes, where the ion maintains the same coordination shell, and \emph{solvent exchange}, where a previously coordinating solvent molecule leaves or a new one enters the solvation shell. In polymers, events include \emph{intrachain} hops, corresponding to a change in the atom index of the coordinating polymer atom within the same polymer chain, and \emph{interchain} hops, corresponding to coordination transfer between distinct polymer chains. In crystalline materials, they comprise \emph{single-ion} and \emph{concerted} hops, depending on whether the motion occurs independently or cooperatively with neighboring ions. These simplified event types are illustrated schematically as $\alpha$, $\beta$, $\gamma$ in Fig.~\ref{fig:overview}c, with detailed definitions provided in the respective sections for liquids, polymers, and crystals. For each ion and each time window, every transition from $S_i^t$ to $S_i^{t{+}\Delta t}$ is assigned to one event type and associated with its corresponding displacement $\mathbf{d}_i(t,\Delta t)$.

\paragraph*{Defining macrostates and transitions between them.}
While transitions between microstates provide insight into how transport occurs through atoms entering or leaving the coordination shell, they do not indicate the type of coordination environment present before or after the transition. A complementary approach to capture this information is to define \emph{macrostates}, where each \emph{macrostate} $\mathbb{S}_i^t$, denoted as X, Y, or Z in Fig.\ref{fig:overview}\textbf{c}, is a discrete coarse-grained label obtained by grouping $S_i^t$ into chemically interpretable categories, thereby anonymizing the specific atom or molecule indices of the coordinating species. Examples include lithium solvated as SSIP, CIP, or AGG in liquids; lithium in activated vs.\ non-activated environments in crystals; and lithium bound by one chain, multiple chains, or no chains in polymers. The windowed sequence ${S_i^t}$ is treated as a discrete stochastic process. Similar to microstate transitions, each transition from $\mathbb{S}_i^t$ to $\mathbb{S}_i^{t{+}\Delta t}$ and its corresponding displacement $\mathbf{d}_i(t,\Delta t)$ is mapped onto one of the pairwise transitions in the macrostate transition diagram between the predefined set of macrostates relevant to the system, as illustrated in the lower part of Fig.\ref{fig:overview}\textbf{c}. The macrostate transition diagram additionally informs the equilibrium populations of macrostates and the transition probabilities \(p_{P\rightarrow Q}(\Delta t)\) between them, where $P$ and $Q$ denote the source and destination macrostates, respectively.

\begin{figure*}[!ht]
\includegraphics[width=0.95\textwidth]{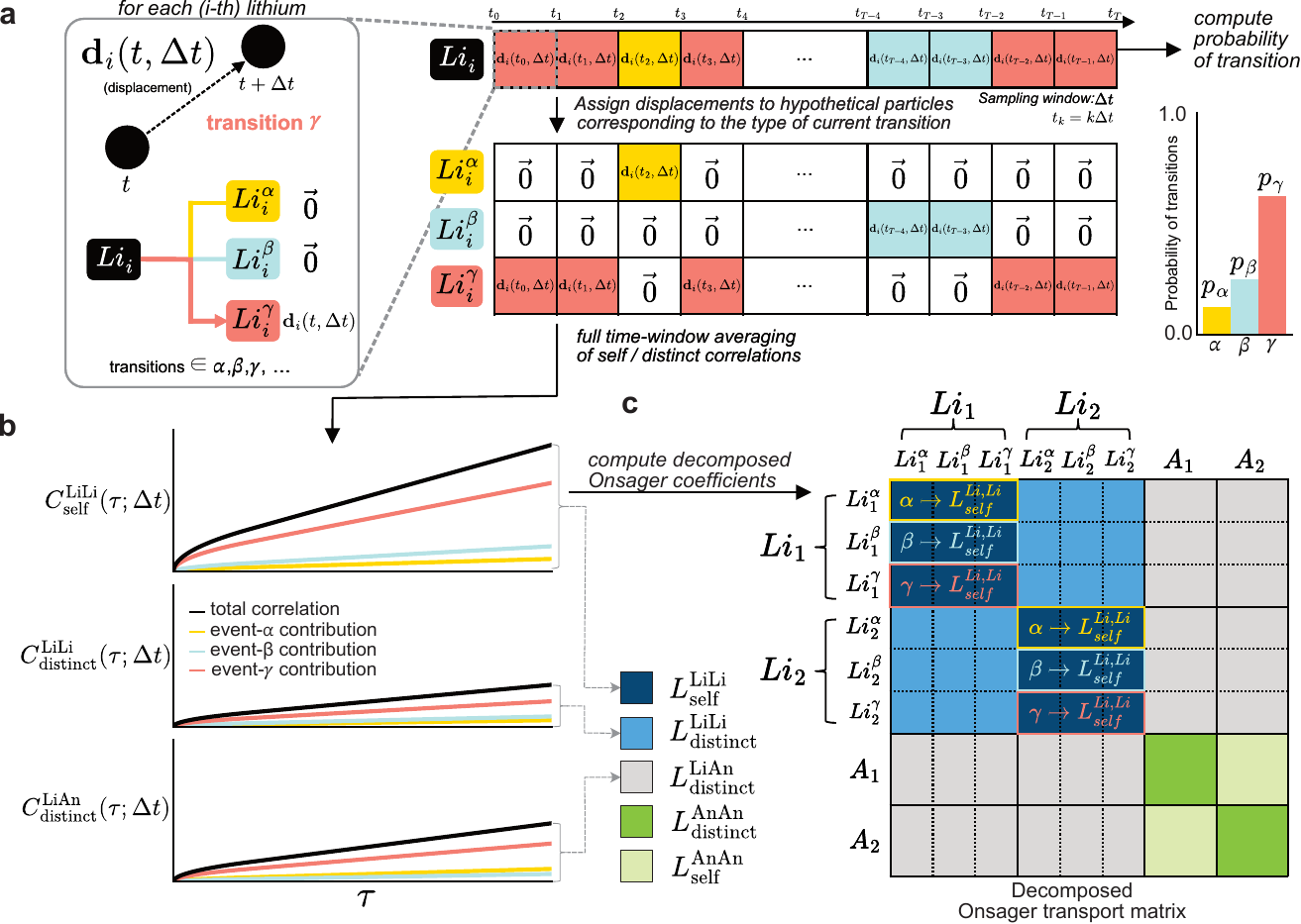}
\caption{
\textbf{Exact additive decomposition of Onsager transport coefficients via the event-specific virtual particle construction.}
\textbf{a}, For each ion \(i\) and sampling window \(\Delta t\), the event indicator $\chi_i^m(t,\Delta t)$ marks whether the microstate transition \(S_i(t)\!\to\!S_i(t{+}\Delta t)\) belongs to event class \(m\) (schematically \(\alpha,\beta,\gamma\)). The event-specific virtual particle displacement illustrates that each virtual particle undergoes displacement only when its parent particle experiences the specific event $m$, such that exactly one class receives the ion’s displacement in each window. In the schematic, white boxes denote \(\vec{0}\) and colored boxes denote \(\mathbf{d}_i(t,\Delta t)\) for the active event in each window. Aggregating information  over all ions \(i\) yields the event probabilities.
\textbf{b}, Displacement correlations at lag \(\tau\) for a given sampling window \(\Delta t\). The displacement correlations associated with Li: $C_\mathrm{self}^{\mathrm{LiLi}}$, $C_\mathrm{distinct}^{\mathrm{LiLi}}$, $C_\mathrm{distinct}^{\mathrm{LiAn}}$ (black lines) equals the sum of event-resolved terms (colored lines) at all $\tau$ values, confirming exact additivity. The slopes of $C(\tau; \Delta t)$ with respect to $\tau$ at long times yield the decomposed Onsager coefficients, and thus the additivity of $C$ directly translates to the additivity of these decomposed slopes.
\textbf{c}, Schematic illustration of the decomposed Onsager transport matrix for a system with two Li and two anions. Solid lines indicate components of the conventional Onsager matrix, whereas the dotted lines indicate the event-decomposed one. Boxes labeled \(\alpha,\beta,\gamma\) denote contributions from individual event types \(\alpha,\beta,\gamma\) to \(L_{\mathrm{self}}^{\mathrm{LiLi}}\). Analogous decompositions apply to \(L_{\mathrm{distinct}}^{\mathrm{LiLi}}\) and \(L_{\mathrm{distinct}}^{\mathrm{Li,Anion}}\).
}
\label{fig:onsager_decomposition}
\end{figure*}

\paragraph*{Virtual particle construction for additivity.}
To enable additive decomposition of Onsager transport coefficients, we introduce the event indicator $\chi_i^m(t,\Delta t)$, which specifies whether ion $i$ at time $t$ undergoes event type $m$ within the window of length $\Delta t$:
\begin{equation}\label{eq:event indicator}
\chi_i^m(t,\Delta t)=
\begin{cases}
1,& \text{if window }[t,\,t{+}\Delta t]\text{ for ion }i\text{ is}\\
&\text{classified as event }m\\
0,& \text{otherwise}
\end{cases}
\end{equation}
so that $\sum_m \chi_i^m(t,\Delta t)=1$, ensuring a disjoint and complete classification of time windows. Here, event type $m$ can represent either microstate or macrostate transitions. In this construction, each lithium ion is represented by $m$ \emph{virtual particles}, each corresponding to one event type. A given virtual particle undergoes displacement only during the windows in which its associated event occurs for the parent ion (further details provided later in this section). The corresponding event-specific virtual particle displacement is defined as following.
\begin{equation}
\label{eq:virtualparticle}
\mathbf{d}_i^m(t,\Delta t)=\chi_i^m(t,\Delta t)\mathbf{d}_i(t,\Delta t)
\end{equation}

We denote $\tau$ as the \emph{lag time} entering correlation functions. Given an MD trajectory analyzed with sampling interval $\Delta t$, $\tau$ must be an integer multiple of $\Delta t$ ($\tau=n\Delta t$, $n\in\mathbb{N}$). As such, the displacement of ion $i$ over lag time $\tau$, conditioned on this sampling interval (denoted by $|\Delta t$) can be expressed as a sum of $n$ adjacent window displacements of length $\Delta t$:
\begin{equation}
\mathbf{d}_i(t, \tau| \Delta t) = \sum_{k=0}^{n-1} \mathbf{d}_i(t + k\Delta t, \Delta t)
\end{equation}
Similarly, the event-specific displacement of $m$-th virtual particle for particle index $i$ over $\tau$ is

\begin{equation}
\mathbf{d}_i^m(t, \tau| \Delta t) = \sum_{k=0}^{n-1} \mathbf{d}_i^m(t + k\Delta t, \Delta t)
\end{equation}

For species $A$ and $B$ (e.g., Li and anion), the conventional time-window-averaged displacement correlation at lag $\tau$ given a trajectory sampled every $\Delta t$ is
\begin{equation}
\label{eq:corr}
C^{AB}(\tau|\Delta t)=\Big\langle\sum_{i\in A}\mathbf{d}_i(t,\tau| \Delta t)\cdot\sum_{j\in B}\mathbf{d}_j(t,\tau| \Delta t)\Big\rangle_t,
\end{equation}
where $\langle\cdot\rangle_t$ denotes averaging over all window start times $t$. The event-resolved counterpart attributes the ion-type $A$’s displacements to event $m$:
\begin{equation}
\label{eq:corr_m}
C^{AB,m}(\tau|\Delta t)=\Big\langle\sum_{i\in A}\mathbf{d}_i^m(t,\tau| \Delta t)\cdot\sum_{j\in B}\mathbf{d}_j(t,\tau| \Delta t)\Big\rangle_t.
\end{equation}
Because ${\chi_i^m}$ is a disjoint, complete partition, we obtain the \emph{additive decomposition} of displacement correlations.
\begin{equation}
\label{eq:additivity}
C^{AB}(\tau|\Delta t)=\sum_m C^{AB,m}(\tau|\Delta t),
\end{equation}

Furthermore, correlations between same particle types ($C^{AA}(\tau|\Delta t)$) can be expressed as a sum of self and distinct correlations following
\begin{equation}
C^{AA,m}_{\mathrm{self}}(\tau| \Delta t)
= \Big\langle \sum_{i \in A}
\mathbf{d}_i^m(t,\tau| \Delta t) \cdot \mathbf{d}_i(t,\tau| \Delta t)
\Big\rangle_t
\end{equation}
\begin{equation}
C^{AA,m}_{\mathrm{distinct}}(\tau| \Delta t)
= \Big\langle \sum_{\substack{i,j \in A \\ i \ne j}}
\mathbf{d}_i^m(t,\tau| \Delta t) \cdot \mathbf{d}_j(t,\tau| \Delta t)
\Big\rangle_t
\end{equation}
such that Equation~\ref{eq:additivity} holds for both the self and distinct terms.

\paragraph*{From correlations to Onsager coefficients.}
Onsager transport coefficients \(L^{AB}\) provide a general, species-resolved description of correlated ionic motion based on linear nonequilibrium thermodynamics\citep{onsager1930reciprocal-257, onsager1931reciprocal-901}. In the conventional Onsager transport matrix (solid lines in Fig.\ref{fig:onsager_decomposition}\textbf{c}), each matrix element quantifies how displacements of species \(A\) correlate with those of species \(B\)\citep{fong2019ion-64a, FongPerssonMacromolecules2020, FongPerssonMacromolecules2021, wheeler2004molecular-e70, ZhouMillerTheJournalofPhysicalChemistry1996} where $A$ and $B$ can represent Li and the anion. Diagonal elements such as $L^{AA}$ can be separated into a \emph{self} term \(L^{AA}_{\mathrm{self}}\) and a \emph{distinct} term \(L^{AA}_{\mathrm{distinct}}\), while off-diagonal elements \(L^{AB}\) with \(A{\neq}B\) encode cross-species correlations (e.g., cation–anion).

The Onsager matrix can be obtained from equilibrium MD trajectories via the Green–Kubo formalism using velocity correlations or, equivalently, via the Einstein formulation from displacement–displacement correlations. We adopt the latter to explicitly retain information of displacements within each $\Delta t$ window as follows:
\begin{equation}
\label{eq:einstein}
L^{AB}=\frac{1}{6Vk_BT}\,\lim_{\tau\to\infty}\,\frac{1}{\tau}\,C^{AB}(\tau|\Delta t),
\end{equation}
where $T$, $V$, $k_B$ denote the temperature, volume, and Boltzmann constant, respectively. Combining \cref{eq:additivity,eq:einstein} yields the event-resolved decomposition (shown as dotted lines in Fig.\ref{fig:onsager_decomposition}\textbf{c})
\begin{equation}
\label{eq:einstein_decomp}
L^{AB}=\sum_m L^{AB,m}
\end{equation}
where $m$ represents either predefined microstate and transitions between macrostate (Fig.\ref{fig:overview}\textbf{c}).
Equation\ref{eq:einstein_decomp} can be applied to both self and distinct components in the case of $L^{AA}$ of same particle type correlation. 

Onsager transport coefficients connect directly to experimental observables such as the Li self-diffusion coefficient ($D_{\mathrm{self}}^{\mathrm{Li}}$) from PFG-NMR, total Li conductivity ($\sigma_{\mathrm{total}}^{\mathrm{Li}}$) from EIS, and Nernst-Einstein Li conductivity ($\sigma_{\mathrm{NE}}^{\mathrm{Li}}$) through the following relations:
\begin{equation}
\begin{aligned}
D^{\mathrm{Li}}_{\mathrm{self}} &= \frac{k_BT}{c_{\mathrm{Li}}}\,L^{\mathrm{LiLi}}_{\mathrm{self}},\qquad
\sigma_{\mathrm{NE}}^{\mathrm{Li}} = F^2\,z_+^2\,L^{\mathrm{LiLi}}_{\mathrm{self}},\\
\sigma_{\mathrm{total}}^{\mathrm{Li}} &= F^2\!\left[z_+^2\!\left(L^{\mathrm{LiLi}}_{\mathrm{self}}+L^{\mathrm{LiLi}}_{\mathrm{distinct}}\right)
+ z_+ z_-\,L^{\mathrm{LiAn}}_{\mathrm{distinct}}\right]
\end{aligned}
\end{equation}
where \(c_\mathrm{Li}\) is the Li concentration, \(z_\pm\) are the ionic charges, $k_B$ is Boltzmann constant, $T$ is temperature, \(\sigma_{\mathrm{NE}}^{\mathrm{Li}}\) is the Nernst–Einstein conductivity (self terms only) and \(\sigma\) is the total conductivity of the Onsager matrix that includes different correlations, often referred to as Wheeler-Newman conductivity\citep{wheeler2004molecular-e70}. In practice, distinct correlations typically require substantially longer trajectories or larger cells to converge than self terms. Therefore, in this work, we primarily focus on decomposing \(\sigma_{\mathrm{NE}}^{\mathrm{Li}}\) and $D^{\mathrm{Li}}_{\mathrm{self}}$ into event-resolved contributions and report exemplary decomposition of cross terms in the Supplementary Information Figure\ref{fig:si_correlations}.

\paragraph*{Probability and effectiveness of transitions.}
The probability of occurrence of event $m$ at a sampling frequency of \(\Delta t\) is
\begin{equation}
\label{eq:prob}
p_m(\Delta t)=\frac{\text{number of windows labeled }m}{\text{total number of $\Delta t$ windows}}
\end{equation}
averaged over all Li particles.
We define the \emph{effectiveness} of \(m\) transport mode as the event-specific decomposed transport coefficient normalized by its probability of occurrence ($\frac{D_{\mathrm{self}}^{\mathrm{Li},m}(\Delta t)}{p_m(\Delta t)}$). Conventional $D_{\mathrm{self}}^{\mathrm{Li}}$ can be retrieved as the average of effective $D_{\mathrm{self}}^{\mathrm{Li}}$ weighted by its probability. This allows us to understand which types of modes are most effective in overall transport and identify the limiting step.

\paragraph*{Scanning across sampling windows.}
Distinct chemical processes (or events) each possess characteristic timescales. In this framework, the sampling frequency serves as a probe to detect distinct types of events that manifest at different characteristic times. By systematically scanning across a wide range of $\Delta t$, we capture how the contribution, frequency, and effectiveness of various event types evolve with timescale. Decomposition of transport coefficients at small $\Delta t$ highlights fast local motions such as vibrations and sub-picosecond exchanges, whereas intermediate $\Delta t$ values resolve mechanisms according to their intrinsic timescales. At very large $\Delta t$, multiple events are coalesced within a single window, such that different physical phenomena become temporally averaged and the ability to distinguish them is lost.

\paragraph*{Computational aspects.}
Accurate decomposition imposes two practical requirements: (i) trajectories that are sufficiently long for even the rarest event type $m$ to be sampled enough for the corresponding \(L^{AB,m}\) to converge (i.e.to reach linearity between \(C^{AB,m}(\tau)\) and \(\tau\)), and (ii) a trajectory saving frequency (intervals at which atomic coordinates are recorded) short enough to resolve the fastest transitions of interest. These requirements entail computing displacement correlations over many overlapping windows for all particle pairs, evaluating \(C^{AB}(\tau;\Delta t)\) for all start times \(t\) and lags \(\tau\). A naive full sliding-window implementation scales as \(\mathcal{O}(T^2)\) with the number of stored frames \(T\). To address this, we instead implement fast Fourier transform (FFT)-based algorithm that reduces this computational cost to \(\mathcal{O}(T\log T)\) for evaluating sliding-window average (see Supplementary Information \ref{si:fft_computation}).

In the sections that follow, we demonstrate \textit{OnsagerDecomposer} on three prototypical lithium-ion conductors: an inorganic crystalline electrolyte, a polymer electrolyte, and a liquid electrolyte, each hosting long-debated transport mechanisms. For each materials class, we formalize the set of microstate transitions as well as macrostate definitions to elucidate prevailing hypotheses (e.g., single-ion versus concerted motion in crystals; interchain, intrachain, and vehicular motion in polymers; vehicular versus solvent/anion exchange in liquids). We then quantify, as functions of \(\Delta t\), the \emph{frequency} and \emph{effectiveness} of each process and extract event-resolved activation energies. This yields a mechanism-resolved, additive decomposition of the Onsager coefficients across different materials classes, providing a quantitative understanding of previously proposed transport mechanisms and revealing the characteristic temporal scales at which distinct mechanisms operate, thereby offering overarching rules for achieving fast ion transport.

\subsection{Concerted hop in inorganic lithium-ion conductors}

In inorganic crystalline solid electrolytes, it has been reported that lithium-ion hops exhibit significant spatiotemporal correlations, often referred to as \emph{concerted hops}, where nearby lithium ions undergo hops together within a short time window\citep{HeMoNatureCommunications2017}. While concerted hops have been observed in MD trajectories through visual inspection\citep{lopez2024how-859, HeMoNatureCommunications2017, JalemKanamuraChemistryofMaterials2013}, and activation energies computed using NEB calculations for concerted hops have been shown to be lower than that of single-ion hops\citep{HeMoNatureCommunications2017, XiaoCederAdvancedEnergyMaterials2021}, three quantitative gaps remain: how much of the macroscopic transport is contributed by concerted versus single-ion motion, what is the spatiotemporal scale that distinguishes single-ion versus concerted hop, and how does Li-stuffing modulate both the frequency and effectiveness of these transport modes. Here, we demonstrate how the framework closes these gaps by providing a mechanism-resolved decomposition of \(D^{\mathrm{Li}}_{\mathrm{self}}\) and event-resolved activation energies across sampling windows \(\Delta t\) and temperature.

\begin{figure*}[!ht]
\includegraphics[width=1.0\textwidth]{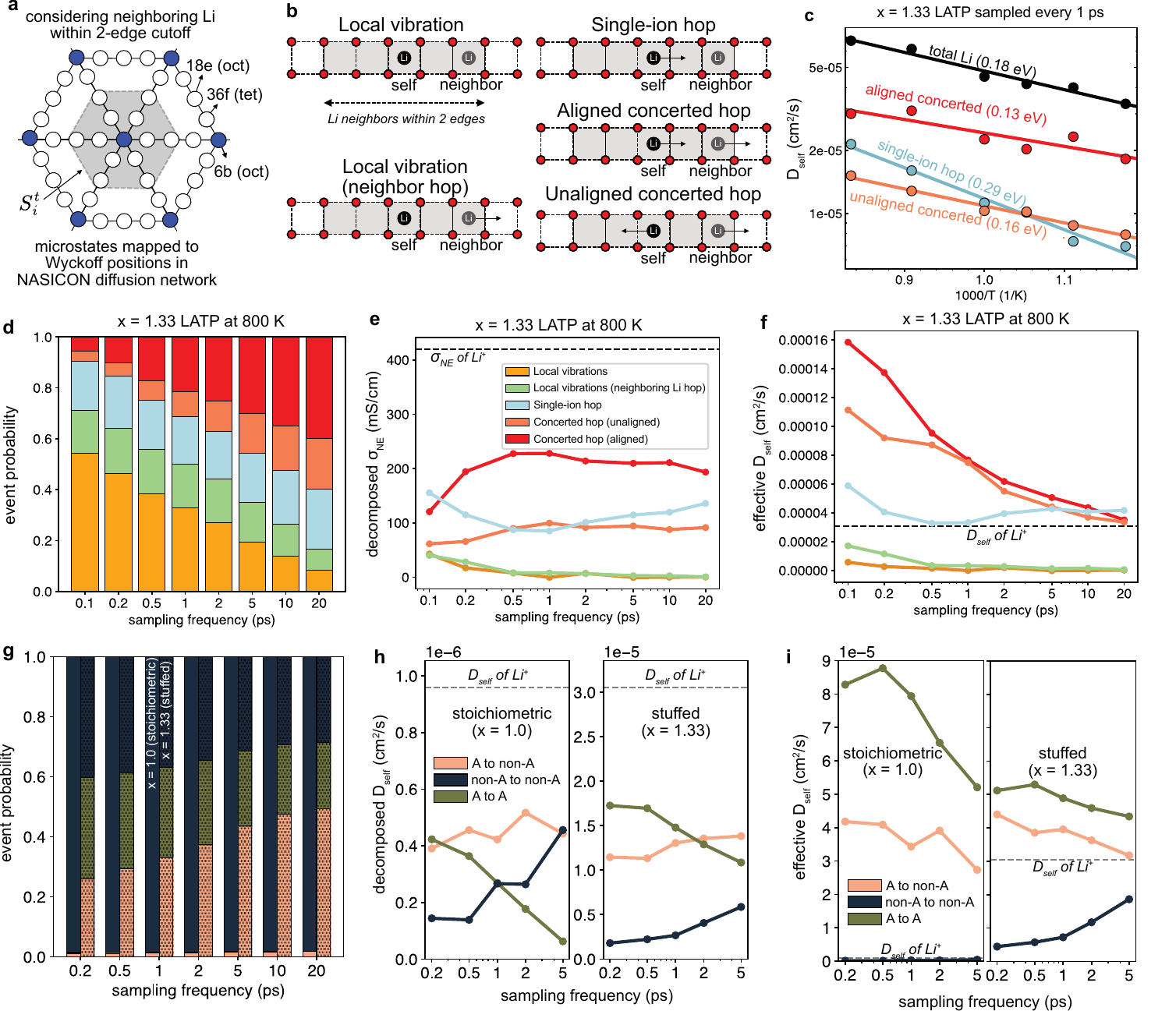}
\caption{
\textbf{Mechanistic decomposition of transport in a crystalline solid electrolyte.}
\textbf{a}, Diffusion network for \ce{Li_{1+x}Al_xTi_{2-x}(PO4)3} (LATP): nodes are Li Wyckoff sites (\(6b\), \(18e\), \(36f\)); edges indicate face-sharing connectivity between polyhedra. Simultaneous hops are detected within a two-edge neighborhood.
\textbf{b}, Event classification for transitions between coordination microstates $S_i^t\!\to\!S_i^{t{+}\Delta t}$: (i) local vibration (no hop; no neighbor hop), (ii) local vibration with neighbor hop, (iii) single-ion hop, (iv) aligned concerted hop (at least one neighbor hops with inter-hop angle \(\le \theta_{\mathrm{align}}\), here \(90^\circ\)), and (v) unaligned concerted hop (angle \(> \theta_{\mathrm{align}}\)). Each window receives exactly one label.
\textbf{c}, Arrhenius plot of \(D^{\mathrm{Li}}_{\mathrm{self}}\) (black) and its mechanism-resolved components (colors) for \(x=1/3\) (\ce{Li_{1.33}Al_{0.33}Ti_{1.67}(PO4)3}). Colored lines sum to the black line by construction.
\textbf{d}, Probability of events \(p_m(\Delta t)\) versus sampling window \(\Delta t\) at 800 K.
\textbf{e}, Contributions of the various microstate transition types to \(D^{\mathrm{Li}}_{\mathrm{self}}\) as a function of \(\Delta t\). The sum over events at each \(\Delta t\) equals the undecomposed \(D^{\mathrm{Li}}_{\mathrm{self}}\).
\textbf{f}, Effective self-diffusion for each event, defined as \(D^{\mathrm{Li}}_{\mathrm{self},m}/p_m(\Delta t)\), as a function of \(\Delta t\).
Macrostate-based decomposition of transport in stoichiometric and Li-stuffed NASICONs are shown in \textbf{g-i}.
\textbf{g}, Event (transition) probabilities between activated (A) and non-activated (non-A) states as a function of sampling window \(\Delta t\) for stoichiometric (\(x=0\)) (left) and Li-stuffed (\(x=0.333\)) (right) \ce{Li_{1+x}Al_xTi_{2-x}(PO4)3}.  
\textbf{h}, Contributions of macrostate transitions (\(A{\to}A\), \(A{\leftrightarrow}\mathrm{non}\text{-}A\), \(\mathrm{non}\text{-}A{\to}\mathrm{non}\text{-}A\)) to \(D^{\mathrm{Li}}_{\mathrm{self}}\).
\textbf{i}, Effective $D_{\mathbf{self}}^{\mathrm{Li}}$ for each macrostate transition, defined as its contribution to \(D^{\mathrm{Li}}_{\mathrm{self}}\) normalized by the corresponding event probability.
}

\label{fig:sic_electrolyte}
\end{figure*}

We examined the NASICON-type superionic conductor\citep{AonoAdachiSolidStateIonics1990, aono1989ionic-6b2} \ce{Li_{1+x}Al_xTi_{2-x}(PO4)3} (LATP), whose three-dimensional diffusion network connects Li Wyckoff sites \(6b\) (octahedral), \(18e\) (octahedral), and \(36f\) (tetrahedral) via face-sharing connectivity (Fig.~\ref{fig:sic_electrolyte}\textbf{a})\citep{XiaoCederAdvancedEnergyMaterials2021}. We begin by defining a lithium's coordination microstate at time t, $S^i_t$, as the index of the Wyckoff-position it occupies within the supercell. We further expand $S^i_t$ to track whether other lithium-ions are within two edges on the diffusion network to detect multi-ion motion (Fig.\ref{fig:sic_electrolyte}\textbf{a}). A \emph{hop} is defined as a change of occupied node between \(t\) and \(t{+}\Delta t\) for a given Li. For any pair of simultaneous hops of lithium neighbors, we compute the angle between hop-direction vectors to separate \emph{aligned} and \emph{unaligned} concerted motion. For each Li and window, its coordination microstate transition \(S_i(t)\!\to\!S_i(t{+}\Delta t)\) is assigned to exactly one of five complete, non-overlapping event classes (Fig.~\ref{fig:sic_electrolyte}\textbf{b}): local vibration (self non-hopping, neighbor non-hopping); local vibration with neighbor hop (self non-hopping, neighbor hopping); single-ion hop (self hopping, neighbor non-hopping); aligned concerted hop (self and neighbor hop, angle $\leq 90^\circ$); and unaligned concerted hop (self and neighbor hop, angle $> 90^\circ$).

Figure~\ref{fig:sic_electrolyte}\textbf{d} plots the event frequencies of distinct types of events as a function of sampling frequency for lithium-stuffed (\(x=0.33\)) LATP at 800 K. At a sampling frequency, $\Delta t$, of 0.1 ps, local vibrations (orange and green) are the most probable event, and any type of concerted motion occurs with a probability below 0.1. As the sampling window becomes longer, the frequency of local vibrations diminishes, compensated by the increasing probability of concerted hops, whereas the frequency of single-ion hops remains roughly constant. This trend arises because once the sampling frequency exceeds the characteristic timescale of single-ion or concerted hops, most windows begin to capture at least one hopping event and are therefore no longer classified as local vibrations. Simultaneously, multiple distinct events with shorter characteristic timescales can occur within a single window, making concerted motions more probable.

Figure~\ref{fig:sic_electrolyte}\textbf{e} decomposes \(\sigma_{\mathrm{NE}}^{\mathrm{Li}}\) into contributions from the five types of microstate transitions. In this plot, the sum of the contributions from the five events mathematically equals the total \(\sigma_{\mathrm{NE}}\) shown as the dashed line. We observe that although the probability of local vibrations exceeds 70\% even at a sampling frequency of 1 ps, their net contribution to \(\sigma_{\mathrm{NE}}\) is nearly zero. In principle, there should be exactly zero contributions from local vibration events in crystalline solid electrolytes due to the immobile framework (non-Li atoms). At a sampling frequency of 0.1 ps, however, we observe a non-zero contribution from local vibrations because such a high frequency misclassifies part of a hopping motion as local vibrations. As the sampling frequency increases beyond 0.5 ps, these contributions diminish completely to zero. The largest contribution consistently comes from aligned concerted hops across all sampling frequencies beyond 0.1 ps. At a sampling frequency of 1 ps, 75\% of \(\sigma_{\mathrm{NE}}^{\mathrm{Li}}\) is attributed to aligned or unaligned concerted hops, and 25\% to single-ion hops.

In addition, by normalizing the decomposed \(D_{\mathrm{self}}\) of each event type by their frequencies, we compute the effective self-diffusion coefficient, as plotted in Figure~\ref{fig:sic_electrolyte}\textbf{f}. By construction, the total \(D_{\mathrm{self}}\) of lithium (black dashed line) can be retrieved as the weighted sum of the effective diffusion coefficients, with the weights given by their corresponding probabilities. We find that, for sampling frequencies beyond 0.2 ps, the effective \(D_{\mathrm{self}}\) of local vibration events diminishes to zero, as expected for crystalline inorganic conductors. At sampling frequencies between 0.1 and 2 ps, concerted hops are more than twice as effective as single-ion hops. Aligned concerted hops are also more effective than unaligned concerted hops. When the sampling frequency exceeds 5 ps, it becomes practically impossible to distinguish concerted hops from single-ion hops in Figure~\ref{fig:sic_electrolyte}\textbf{f}, suggesting that such temporal scales surpass the characteristic timescales\citep{JunCederProceedingsoftheNationalAcademyofSciences2024} of both mechanisms.

Based on the analysis above, we take 1 ps as the representative timescale to distinguish the five types of events in Li–NASICONs, and use this timescale to decompose the temperature-dependent self-diffusion coefficients, as shown in Figure~\ref{fig:sic_electrolyte}\textbf{c}. This enables us to compute activation energies for distinct transport modes. We find that the activation energy of aligned concerted hops (0.13 eV) is significantly lower than that of single-ion hops (0.29 eV). This result is in excellent agreement with microscopic models from NEB calculations, which indicate that the concerted hop process in Li–NASICONs exhibits an activation energy of 0.19 eV\citep{LangElsasserChemistryofMaterials, XiaoCederAdvancedEnergyMaterials2021}. To our knowledge, this is the first time a process-dependent activation energy has been quantitatively extracted from MD simulations.

A widely accepted picture for superionic crystals (NASICONs, garnets, thiophosphates) is that “lithium stuffing” via aliovalent doping (e.g., \ce{Ti^{4+} -> Al^{3+}}) raises local Li site energies through Li–Li repulsion, thereby creating \emph{activated} local Li environments that lower migration barriers and preferentially enable \emph{concerted} multi-ion motion over independent single-ion hops\citep{XiaoCederAdvancedEnergyMaterials2021}. To quantitatively connect concerted motion with transitions between activated and non-activated environments, we aggregate microstates into two macrostates: activated (A), when a lithium has another lithium within a two-edge distance in the diffusion network, and non-activated (non-A), when no lithium is present within that range. Macrostate A corresponds to a high-energy state due to Li–Li Coulombic repulsion. As such activated environments are known to arise from lithium stuffing, we compare stoichiometric (\(x=0\)) and Li-stuffed (\(x=0.333\)) Li-NASICONs at 800 K.  

Figure~\ref{fig:sic_electrolyte}\textbf{g} reports the transition probabilities between A and non-A states across a range of sampling frequencies. A 33\% increase in Li concentration (from 1 to 1.33 per formula unit) shifts the system from exhibiting nearly no A→A transitions to a regime where most transitions involve the A-state. This can be explained by the fact that the ground state of stoichiometric NASICON (Figure~\ref{fig:sic_electrolyte}\textbf{a}) contains no A-state; therefore, Li transport must proceed through a sequential process of A-state formation and annihilation (i.e., A\(\rightarrow\)non-A transitions). In contrast, the ground state of stuffed NASICON already contains a mixture of A-state and non-A-state Li, with the ratio determined by the stuffing level \(x\). Figure~\ref{fig:sic_electrolyte}\textbf{h} decomposes the corresponding Li self-diffusion coefficient into contributions from macrostate transitions. In the stuffed case, the majority of \(D_{\mathrm{self}}\) are attributed to A→A and A→non-A transitions, whereas in the stoichiometric material, most of \(D_{\mathrm{self}}\) is carried by A→non-A transitions. 

Figure~\ref{fig:sic_electrolyte}\textbf{i} shows the \emph{effective} \(D_{\mathrm{self}}\) for each transition type, obtained by normalizing their contributions by event frequency. In the stoichiometric material, the effective \(D_{\mathrm{self}}\) of non-A→non-A transitions are negligible, and in the stuffed case they remain the least effective mode. In contrast, transitions involving activated lithium environments (A→A and A→non-A) exhibit comparable effectiveness in both stoichiometric and stuffed compositions. While \(D_{\mathrm{self}}\) is limited by the most frequent but ineffective non-A→non-A transitions, the stuffed system achieves higher conductivity by more frequently accessing effective modes of transport. These results, as well as the microstate decomposition result of stoichiometric NASICON in Supplementary Figure\ref{fig:si_stoichiometric_ltp} demonstrate that the origin of superionic conductivity, and the orders-of-magnitude improvement in Li-stuffed NASICONs, arises from the increased population of activated local environments from a small amount of excess Li that enables highly effective transitions involving the high-energy A-state. Concerted hops, which can only be accessed through activated environments, emerge as the dominant and most effective mechanism in Li-stuffed crystalline conductors. The characteristic timescale of these concerted processes lies at around 1 ps, where their effectiveness is most clearly distinguished from that of single-ion hops with negligible contribution from local vibration events.

\subsection{Polymer electrolytes: vehicular motion following segmental dynamics vs.\ inter/intrachain hopping}

\begin{figure*}[!ht]
\includegraphics[width=1.0\textwidth]{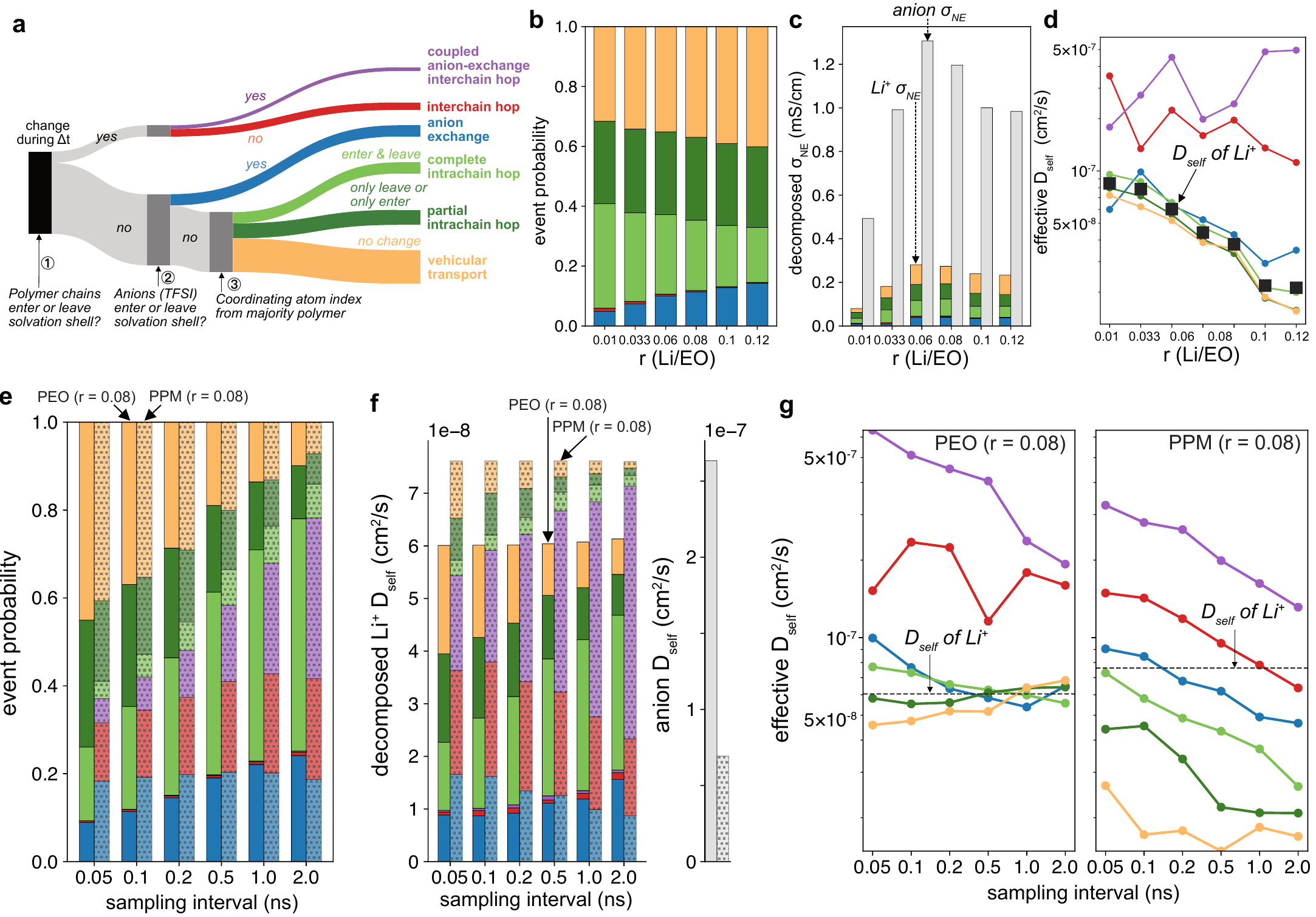}
\caption{
\textbf{Mechanistic decomposition of transport in polymer electrolytes.}
\textbf{a}, Hierarchical event classification for transitions between coordination microstates in polymer electrolytes, yielding six transition types: vehicular (orange), partial intrachain hop (dark green), complete intrachain hop (light green), interchain hop (red), anion exchange (blue), and \emph{coupled anion-exchange and interchain hop} (purple). These colors are used consistently throughout the figure. 
\textbf{b}, Event probabilities sampled every 200~ps as a function of Li/EO ratio $r$ in PEO. 
\textbf{c}, Decomposed \(\sigma_{\mathrm{NE}}^{\mathrm{Li}}\) (stacked colored bars) and \(\sigma_{\mathrm{NE}}^{\mathrm{anion}}\) (grey bars) as a function of LiTFSI concentration, sampled every 200~ps. 
\textbf{d}, Effective \(\sigma_{\mathrm{NE}}^{\mathrm{Li}}\) (colored points) as a function of $r$. Black squares represent the total \(\sigma_{\mathrm{NE}}^{\mathrm{Li}}\), corresponding to the weighted average of the colored points. 
\textbf{e--g}, Quantitative comparison of ion transport in PEO and PPM polymer electrolytes at $r = 0.08$. 
\textbf{e}, Event probabilities in PEO (solid bars) and PPM (dotted bars) as a function of sampling interval. 
\textbf{f}, Decomposed lithium self-diffusion coefficients \(D^{\mathrm{Li}}_{\mathrm{self}}\) for PEO (solid bars) and PPM (dotted bars) versus sampling interval. \(D^{\mathrm{anion}}_{\mathrm{self}}\) for TFSI is shown in grey bars. 
\textbf{g}, Comparison of effective \(D^{\mathrm{Li}}_{\mathrm{self}}\) in PEO (left) and PPM (right). Black dashed lines indicate the overall \(D^{\mathrm{Li}}_{\mathrm{self}}\) in each system, corresponding to the weighted average of the colored points. 
}
\label{fig:peo_electrolyte}
\end{figure*}

Solid polymer electrolytes (SPEs) represent another important class of Li-ion conductors that have been studied for decades\citep{ChooBalsaraProgressinPolymerScience2020, long2016polymer-fb5}. The prototypical system is lithium salts (e.g., \ce{LiBH4}, \ce{LiTFSI}) dissolved in poly(ethylene oxide) (PEO). Understanding Li-ion transport in SPEs has been particularly challenging due to the complexity of disentangling multiple, overlapping transport modes: vehicular motion that follows segmental chain dynamics, intrachain and interchain hopping, and at high salt concentrations, anion-coupled transport. While numerous studies have examined Li-ion transport in polymer electrolytes\citep{borodin2006mechanism-3ef, maitra2007cation-fa0, diddens2010understanding-ba4, borodin2007li-7b2}, the efficiency of each transport mode, their relative frequencies, and the identity of the rate-limiting step remain qualitative. In this section, we demonstrate that \textit{OnsagerDecomposer} can quantitatively disentangle and evaluate these mechanisms of Li transport in a polymer medium. We demonstrate how contributions and effectiveness of different transport modes change as a function of LiTFSI concentration. We further compare the Li transport mechanism in PEO against a recently developed polymer, poly(pentyl malonate) (PPM)\citep{yu2022practical-115, lee2025toward-92e} whose transference number is over 3 times higher albeit a slightly lower total conductivity. To this end, we performed molecular dynamics simulations of PEO (and PPM) with varying concentrations of LiTFSI (see Methods).  

For polymer electrolytes, we define coordination microstates using a convex-hull-based method that avoids the need for a hard distance cutoff to identify the first solvation shell of lithium. These microstates encode which oxygen atoms (from PEO/PPM or TFSI) or nitrogen atoms (from TFSI) are coordinated to each Li ion at every time frame. Following Method~1 in Fig.~\ref{fig:overview}\textbf{c}, transitions between coordination microstates are classified into six event types, summarized in the Sankey diagram (Fig.~\ref{fig:peo_electrolyte}\textbf{a}). The classification proceeds hierarchically. We first check whether at least one polymer chain enters or leaves the solvation shell (interchain hop), and second whether at least one TFSI molecule enters or leaves (anion-exchange). If both occur simultaneously, the event is labeled as coupled anion-exchange and interchain hop. If none of the solvating molecular residue IDs change, we then inspect whether the solvating atom indices from the majority polymer chain change. Here, the “majority chain” is defined as the polymer residue contributing the largest number of coordinating oxygens during its residence time. If the solvating atom indices from this chain remain unchanged, the event is classified as vehicular transport, in which Li moves together with the segmental motion of the majority chain. If they change, the event is labeled as intrachain hop. Intrachain hops are further sub-classified as complete (oxygen atoms from the majority polymer entering and leaving simultaneously, corresponding to a full site-to-site hop) or partial rearrangements.

PEO–LiTFSI exhibits a well-established maximum in ionic conductivity at a salt concentration of $r = 0.08$\citep{lascaud1994phase-294,pesko2017negative-3e6} where $r$ is defined as the ratio of moles of \ce{Li+} to moles of oxygen atoms on the polymer chains. This non-monotonic behavior has been qualitatively attributed to increasing carrier density at higher salt loadings, counteracted by Li–polymer crosslinking\citep{diddens2010understanding-ba4, borodin2006mechanism-3ef} and Li–anion clustering\citep{molinari2018effect-7bb, france-lanord2020effect-522}, which slows ion motion. However, the microscopic processes responsible for these effects have not been quantitatively resolved.

Figure~\ref{fig:peo_electrolyte}\textbf{b–d} show how event frequencies, their contributions to $\sigma_{\mathrm{NE}}^{\mathrm{Li}}$, and their per-event effectiveness evolve with salt concentration in PEO. Intrachain hops and vehicular events dominate the statistics, whereas interchain-hop–related events are already rare at low concentration and vanish entirely at higher salt loadings (Fig.\ref{fig:peo_electrolyte}\textbf{b}). From r = 0.01 to r = 0.08, the largest increases in $\sigma_{\mathrm{NE}}^{\mathrm{Li}}$ (Fig.\ref{fig:peo_electrolyte}\textbf{c}) arise from strengthening contributions of intrachain hops and anion exchange. At concentrations above r = 0.08, $\sigma_{\mathrm{NE}}^{\mathrm{Li}}$ plateaus because both intrachain-hop contributions decline and interchain-hop–related pathways disappear, suppressing the polymer-rearrangement mechanisms that most effectively transport lithium in PEO. This interpretation is supported by the per-event effectiveness $D_{\mathrm{NE}}^{\mathrm{Li}}$ (Fig.\ref{fig:peo_electrolyte}\textbf{d}): at all concentration regimes, interchain-hop–related events are by far the most effective mode of transport, whereas all other events that are coupled to polymer chain segmental dynamics are much less effective and further lose effectiveness with increasing concentration. Among them, complete intrachain hops are moderately effective whereas vehicular events remain the least effective mode of transport across all concentration regime. The overall $D_{\mathrm{self}}^{\mathrm{Li}}$, which matches the weighted average of these contributions, decrease monotonically with concentration, because most of $D_{\mathrm{self}}^{\mathrm{Li}}$ are dominated by segmental dynamics (not interchain hops) and the segmental dynamics slows down with more lithium salts as seen in the self-diffusion coefficient of polymer constitutional repeat unit (CRU) $D_{\mathrm{self}}^{\mathrm{CRU}}$(Supplementary Information \ref{si:correlation_polymer}).

Together, these results provide a quantitative mechanistic explanation for the well-known conductivity maximum in PEO: increasing concentration enhances anion exchange and intrachain hopping up to $r = 0.08$, after which polymer crosslinking and the resulting suppression of intrachain hops dominate. The decomposition identifies which microscopic transitions accelerate or impede ion transport, thereby clarifying how molecular design and salt loading jointly shape conductivity in polymer electrolytes.

Subsequently, we use our algorithms to quantitatively compare transport mechanisms across temporal scales in PEO and PPM. Figure~\ref{fig:peo_electrolyte}\textbf{e} shows the probability of each transition type as a function of sampling window at $r = 0.08$. While interchain-hop–related events are extremely rare in PEO, they are among the most frequent events in PPM, accounting for $\sim$ 15\% of transitions at a 0.05 ns sampling interval and exceeding 60\% at longer intervals. In PEO, vehicular transport is the most frequent event at short timescales, whereas complete intrachain hops become the most probable at longer times; overall, vehicular and intrachain-hop events together account for over 70\% of transitions across all temporal scales. In PPM, by contrast, the probability of intrachain hops remains relatively constant, while the probability of vehicular events decreases with increasing sampling interval and is compensated by a sharp rise in interchain-hop probabilities.

Figure~\ref{fig:peo_electrolyte}\textbf{f} compares the decomposed $\mathrm{D}^{\mathrm{Li}}_{\mathrm{self}}$ across temporal scales. By construction, the sum of the mechanism-resolved components (colored bars) equals the total $\mathrm{D}^{\mathrm{Li}}_{\mathrm{self}}$. Overall, $\mathrm{D}^{\mathrm{Li}}_{\mathrm{self}}$ in PPM is slightly higher than in PEO. However, $\mathrm{D}^{\mathrm{anion}}_{\mathrm{self}}$ in PEO is significantly higher than in PPM. As the ratio $\mathrm{D}^{\mathrm{Li}}_{\mathrm{self}}$ / $\mathrm{D}^{\mathrm{Li}}_{\mathrm{self}}+\mathrm{D}^{\mathrm{anion}}_{\mathrm{self}}$ is equal to the Nernst–Einstein approximation of the cation transference number, our results agree with previous reports showing that PPM exhibits a more than threefold higher transference number than PEO. Inspecting the components of $\mathrm{D}^{\mathrm{Li}}_{\mathrm{self}}$, most Li transport in PEO arises from vehicular and intrachain-hop mechanisms, whereas in PPM, interchain-hop events account for the majority of Li transport.

We further compare the effective $\mathrm{D}^{\mathrm{Li}}_{\mathrm{self}}$ of each mechanism across timescales in PEO and PPM (Fig.~\ref{fig:peo_electrolyte}\textbf{g}). In both materials, interchain-hop–related events are the most effective transport mode on a per-event basis. In PEO, at timescales shorter than 0.5 ns, vehicular transport is the least effective mechanism; at longer times, the effective values of all non–interchain-hop mechanisms converge, reflecting that these timescales exceed the characteristic window needed to distinguish the modes. As these non–interchain-hop modes constitute 99\% of transitions, the overall $\mathrm{D}^{\mathrm{Li}}_{\mathrm{self}}$ in PEO is governed by their effectiveness rather than by the rare interchain hops. In PPM, however, the hierarchical ordering of effectiveness is clear across all timescales: vehicular transport is least effective, followed in increasing order by partial intrachain hops, complete intrachain hops, anion exchange, and (coupled) interchain hops. Compared to PEO, all segmental-dynamics–coupled modes (vehicular and intrachain hops) are significantly less effective in PPM, consistent with the lower $\mathrm{D}^{\mathrm{CRU}}_{\mathrm{self}}$ in PPM (Supplementary Information Figure \ref{fig:correlations_polymer}), indicating slower segmental dynamics. However, the overall $\mathrm{D}^{\mathrm{Li}}_{\mathrm{self}}$ in PPM is strongly dominated by the highly effective interchain-hop mode, which occurs far more frequently in PPM than in PEO.

Our analysis provides a quantitative explanation of the higher $\mathrm{D}^{\mathrm{Li}}_{\mathrm{self}}$ and higher cation transference number in PPM compared with PEO. In PEO, most Li ions undergo vehicular and intrachain-hop motions coupled to Rouse-like polymer dynamics, both of which are slow, and the highly effective interchain-hop events occur too infrequently to compensate. In PPM, polymer segmental dynamics are slower, reducing the effectiveness of vehicular and intrachain-hop modes, yet Li motion is partially decoupled from polymer dynamics and can access frequent and highly effective interchain hops. Meanwhile, the more viscous PPM matrix suppresses TFSI mobility, reducing $\mathrm{D}^{\mathrm{anion}}_{\mathrm{self}}$. Together, these effects explain why PPM exhibits higher $\mathrm{D}^{\mathrm{Li}}_{\mathrm{self}}$ and substantially higher transference number than PEO.

\subsection{Vehicular and anion- and solvent-exchange mechanism in liquid electrolytes}

\begin{figure*}[!ht]
\includegraphics[width=0.95\textwidth]{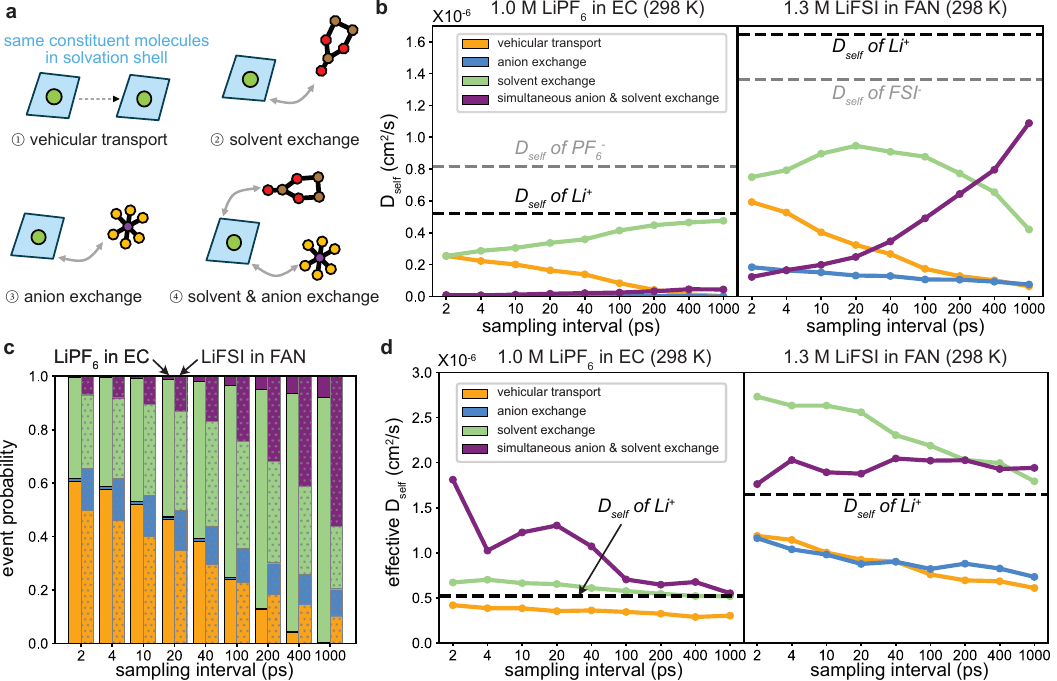}

\caption{
\textbf{Mechanistic decomposition of lithium transport in liquid electrolytes} 
\textbf{a}, Classification of transitions between coordination microstates for liquid electrolytes. If no constituent solvating molecules change, the transition is labeled as \emph{vehicular transport}. If only a solvent molecule enters or leaves the solvation shell, the transition is classified as \emph{solvent exchange}. If only an anion enters or leaves, it is classified as \emph{anion exchange}. If both solvent and anion entities change within the time window, the transition is classified as a \emph{simultaneous solvent–anion exchange} event.  
\textbf{b}, Comparison of decomposed lithium self-diffusion coefficients in EC with 1.0 M \ce{LiPF6} and in FAN with 1.3 M LiFSI. Black and gray dashed lines indicate the total self-diffusion coefficients of lithium and anions, respectively.  
\textbf{c}, Event probabilities as a function of sampling window \(\Delta t\). Solid bars correspond to EC with \ce{LiPF6}, and dotted bars to FAN with LiFSI.  
\textbf{d}, Effective self-diffusion coefficients for the different transition types in EC with \ce{LiPF6} and FAN with LiFSI as a function of \(\Delta t\). The black dashed line shows the total lithium self-diffusion coefficient, which is the weighted average of the colored curves.
}

\label{fig:liquid_electrolyte}
\end{figure*}

Understanding the mechanism of ion transport in liquid electrolytes is critical. In these systems, lithium ions are solvated by sheaths of solvating molecules (solvents and anions) and diffuse through a combination of vehicular transport (motion with fixed solvating entities) and structural transport (resembling a hop). However, the relative timescales and dominance of vehicular versus structural transport have been widely discussed in the field due to the intricacies of interpreting 2D FT-IR results. Some studies, based on two-dimensional FT-IR measurements, have claimed that lithium ions in carbonate electrolytes are transported via a fast solvent-exchange mechanism with their surroundings on the picosecond timescale (also referred to as structural diffusion)\citep{LeeChoNatureCommunications2017}. In contrast, more recent works have argued that solvent-exchange dynamics are negligible below the 100 ps timescale, with most lithium-ion transport instead being driven by vehicular transport\citep{dereka2022exchange-mediated-d09}. Our computational framework can not only resolve the temporal scale of exchange processes but also quantify lithium-ion transport contributions from each mechanism, thereby revealing the effectiveness of these transport modes in liquid electrolytes and providing guidelines for improving Li-ion conductivities.

We demonstrate the mechanistic decomposition of ion transport in a conventional liquid electrolyte by comparing a standard carbonate electrolyte, 1.0 M \ce{LiPF6} in ethylene carbonate (EC) solvent, with a novel formulation, 1.3 M LiFSI in fluoroacetonitrile (FAN) solvent. The latter was recently reported\citep{LuFanNature2024} to exhibit significantly higher ionic conductivity (40.3 mS/cm at 25 $^\circ$C and 11.9 mS/cm at -70 $^\circ$C) enabled by a so-called ligand-channel transport mechanism, in which solvents rapidly exchange to connect distinct solvation configurations. This behavior is reported to be less pronounced in conventional carbonate electrolytes.

For these liquid electrolytes, we define the coordination microstate using a geometric convex hull, keeping track of the residue and atom IDs of all coordinating species in the first solvation shell. Four types of non-overlapping, fully spanning events are defined for liquids in Figure \ref{fig:liquid_electrolyte}\textbf{a}. For a given lithium ion during each time window, if the set of solvating residue IDs of all coordinating molecules does not change, we classify it as vehicular transport. If a solvent entity leaves or enters the solvation shell, we classify it as solvent exchange. Similarly, if an anion entity (e.g., \ce{PF6-} or \ce{FSI-}) leaves or enters the solvation shell, we classify it as anion exchange. If both occur within the same window, we classify it as solvent \& anion exchange. We note that while a chemical exchange process (one-in/one-out) occurring within a short timescale may differ from the sequential loss and subsequent addition of a molecule to the solvation environment with a time lag, our definition of transitions allows us to naturally capture the temporal boundaries of these processes as we scan through a range of the sampling frequencies.

Figure \ref{fig:liquid_electrolyte}\textbf{b, c} compares the decomposed $D_{self}$ and the probability of transitions for 1.0 M \ce{LiPF6} in EC and 1.3 M \ce{LiFSI} in FAN at 298 K. As experimentally reported, the $D_{self}$ of Li is significantly higher in the FAN-based electrolyte compared to the EC-based solvent. For EC, vehicular transport and solvent exchange account for the vast majority of $D_{self}$, while transitions involving anion exchange contribute negligibly to Li transport. When examining contributions as a function of sampling frequency, we find that the fraction from vehicular transport in EC decreases from ~50\% to nearly zero, compensated by an increase in solvent exchange from ~50\% to nearly 100\%. In contrast, in the FAN-based electrolyte we observe a substantial contribution from both solvent exchange and simultaneous solvent–anion exchange processes. Here, as sampling frequency increases, the decrease in vehicular transport is compensated by a rapid increase in simultaneous solvent–anion exchange events.

Examining the probability of transitions as a function of sampling frequency, we find that in the EC-based electrolyte, the probability of vehicular transport exceeds 0.6 at a 2 ps sampling frequency, with the remainder attributed to solvent exchange. However, the vehicular transport probability decreases to below 0.3 at a sampling frequency of 100 ps, corresponding to a rise in solvent exchange probability. Anion exchange events rarely occur without being accompanied by solvent exchange. In the FAN-based electrolyte, the probabilities of anion exchange and solvent exchange exceed 10\% and 20\%, respectively, across all sampling frequencies. The decrease in vehicular transport probability at longer sampling frequencies is compensated by an increase in simultaneous solvent–anion exchange transitions.

A major distinction in the FAN-based electrolyte is that anion-exchange–involving events (purple and blue) occur significantly more frequently than in the EC-based electrolyte across all sampling frequencies. Comparing the solvent-exchange–related transitions (green and purple) of EC and FAN, we find that at all sampling frequencies the solvent exchange probability is higher in EC, indicating that solvent exchange occurs on a faster timescale in EC than in FAN. By contrast, anion-exchange probabilities (blue and purple) are markedly faster in FAN than in EC.

Figure \ref{fig:liquid_electrolyte}\textbf{d} compares the effective $D_{\mathrm{self}}$ of lithium. In the EC-based electrolyte, vehicular transport is the most sluggish mode. The self-diffusivity arises from a balance between the slightly more effective solvent exchange mechanism and the less effective vehicular transport. Anion exchange rarely occurs without simultaneous solvent exchange. Although simultaneous solvent–anion exchange is consistently more effective than the other mechanisms, its occurrence in EC is rare. In contrast, in the FAN-based electrolyte, anion exchange and vehicular transport are the two least effective modes of transport. A notable observation is that solvent exchange is highly effective in this system, particularly on timescales below 100 ps. Simultaneous solvent–anion exchange is similarly effective, though its effectiveness appears to stem primarily from solvent exchange, since anion exchange without solvent involvement is among the least effective mechanisms, along with vehicular transport.

Therefore, our analysis indicates that the origin of the high Li-ion conductivity in the FAN-based electrolyte lies in the significantly more effective solvent exchange mechanism. This is in direct agreement with the proposed ligand-channel mechanism, which describes a transport scenario where the solvation energy of the solvent molecules is weak enough to facilitate rapid solvent exchange, thereby gradually connecting distinct configurations of the solvation shell. Our analysis directly and quantitatively proves that this solvent exchange mechanism in FAN is substantially more effective than EC, and validates the effectiveness of the ligand-channel mechanism. Furthermore, the greater effectiveness of vehicular transport in FAN compared to EC can be rationalized by the significantly lower viscosity of FAN (0.56 mPa$\cdot$s)\citep{LuFanNature2024, RiddickSakano31} relative to EC (1.92 mPa$\cdot$s at 40$^\circ$C)\citep{thompson1982viscosities-cd7}.

\subsection{Coarse-grained macrostates for liquids: transitions between SSIP, CIP, and AGG}

\begin{figure*}[!ht]
\includegraphics[width=0.9\textwidth]{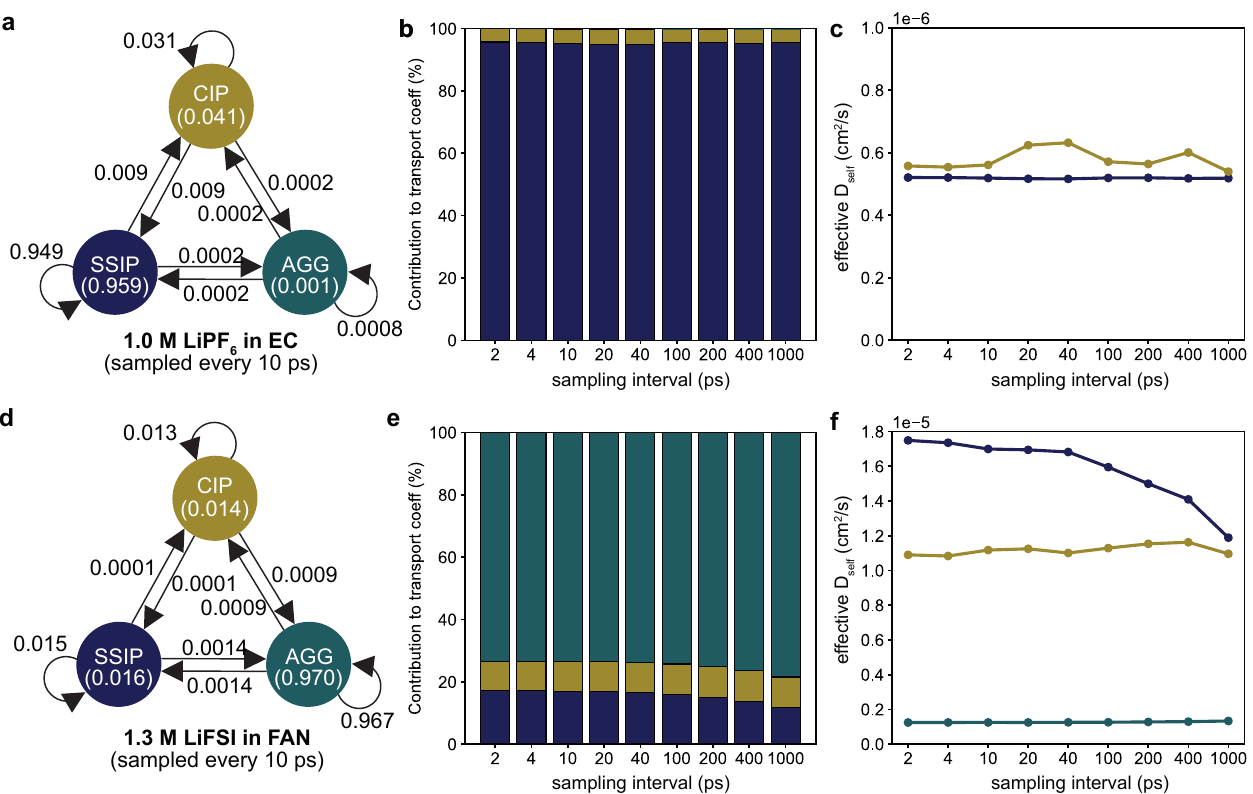}
\caption{
\textbf{Transitions between coarse-grained macrostates reveal the contribution and efficiency of transport across different solvation environments in liquid electrolytes.}
Coarse-grained macrostate diagrams for 1.0 M \ce{LiPF6} in EC (\textbf{a}) and 1.3 M \ce{LiFSI} in FAN (\textbf{d}), where all states are classified by solvation environment into three macrostates: solvent-separated ion pairs (SSIP), contact ion pairs (CIP), and aggregates (AGG). Numbers next to arrows denote transition probabilities, while numbers inside circles indicate the population of macrostates. \textbf{b}, Fractional contributions of $D_{\mathrm{self}}$ and $\sigma_{\mathrm{NE}}$ from events involving SSIP, CIP, and AGG as a function of sampling window for EC; \textbf{e}, corresponding results for FAN. \textbf{c}, Effective $D_{\mathrm{self}}$ as a function of sampling window for EC; \textbf{f}, corresponding results for FAN. AGG is not plotted in \textbf{c} due to large uncertainty from their low population below 0.1\%.
}
\label{fig:reduced_markov}
\end{figure*}

Complementing the insights from microstate transitions, analyzing transitions between macrostates provides an overarching understanding of transport in liquid electrolytes. Solvation environments in these systems are commonly classified as SSIP (no anion species within the solvation shell), CIP (one anion within the solvation shell), and AGG (two or more anions in the solvation shell of a lithium, or at least one anion in the solvation shell that also coordinates another lithium). We assign each lithium's coordination microstate to one of these macrostates (SSIP, CIP, or AGG). Applying our algorithm to this coarse-grained description allows us to determine which types of transitions between macrostates are most effective for transport, and which macrostates themselves exhibit faster dynamics.

Figure \ref{fig:reduced_markov}\textbf{a,d} illustrates the macrostate transition diagrams for EC- and FAN-based electrolytes, respectively. The static populations (shown inside each node) reveal markedly different solvation environments in the two systems. In EC, 95.9\% of lithium ions reside in the SSIP macrostate, 4.1\% in CIP, and AGG is extremely rare, consistent with previous results\citep{hou2021solvation-ec1}. In contrast, in the FAN-based electrolyte, 97\% of lithium ions occupy AGG environments, which is qualitatively consistent with the computational results reported by \citet{LuFanNature2024}. For this analysis, a sampling frequency of 10 ps was used consistently.

Figure \ref{fig:reduced_markov}\textbf{b,e} illustrates the contributions of SSIP, CIP, and AGG to $D_{\mathrm{self}}$. This is done by assigning the decomposed transport coefficient of an X–Y transition equally (0.5 each) to X and Y, where X and Y can be any of the macrostates (SSIP, CIP, or AGG). Such decomposition provides averaged insight into which solvation environments are most effective for ion transport. For the EC-based electrolyte, we find that CIP is slightly more effective than SSIP across all sampling frequencies. Although CIPs are often considered detrimental to Li-ion conduction due to ion-pairing effects\citep{marcus2006ion-42d}, our results provide a new perspective: lithium ions in CIP solvation states are in fact faster. This can be rationalized by the relatively small-sized pairing of a single Li and an anion, which exerts weaker polarization on neighboring solvents and thereby produces less resistance compared to lithium ions unpaired with anions. For EC, the AGG population is too low to yield reasonably converged values.

On the other hand, in the FAN-based electrolyte we observe a striking difference. While the most probable macrostate is AGG, its effective diffusivity is on the same order as that of SSIP and CIP in EC ($\sim10^{-6}$ cm$^{2}$/s). However, the less populated macrostates (SSIP and CIP) are up to an order of magnitude more effective than AGG. In FAN, the slower motion of AGG compared to SSIP and CIP can be rationalized by the large Li--FSI aggregates (see Supplementary Figure\ref{fig:cluster_population} for the cluster population map), which have higher effective mass and therefore diffuse more slowly in the liquid medium than unpaired ions or CIPs. Our computational framework clearly establishes that the enhanced conductivity of FAN arises from the markedly higher effectiveness of CIP and especially SSIP states, which, despite their lower populations, boost the overall Li-ion conductivity in FAN to more than twice that of EC. 

In FAN, we also note that SSIP is more effective than CIP across all observed time windows. Figure \ref{fig:liquid_electrolyte}\textbf{d} shows that solvent exchange is significantly more effective in FAN than anion exchange. SSIP has only FAN molecules in the solvation shell, whereas CIP contains an anion. Consequently, SSIP, with a larger number of nearby FAN molecules, can undergo facile FAN-exchange processes, which may explain its higher effectiveness compared to CIP.

%%%%%%%%%%%%%%%%%%%%%%%%%%%%%%%%%%%%%%%%%%%%%%%%%%%%%%%%%%%%%%%%%%%%%%%%%%%%%%%%
% Discussion
%%%%%%%%%%%%%%%%%%%%%%%%%%%%%%%%%%%%%%%%%%%%%%%%%%%%%%%%%%%%%%%%%%%%%%%%%%%%%%%%

\section{Discussion}
\label{sec:discussion}

The results presented here establish a general and quantitative framework for dissecting ion transport mechanisms across fundamentally different classes of lithium-ion conductors. By introducing definitions of microstates and macrostates that capture the physicochemical nature of solvation, combined with efficient time-window-averaged correlation analysis, we demonstrated that transport in crystalline, polymeric, and liquid electrolytes can be decomposed into distinct mechanistic contributions. This enables, for the first time, a direct quantification of how frequently each type of event occurs, how effective each is at carrying ions, and on what characteristic temporal scales these mechanisms become distinguishable. 

Through our computational framework, we clarify long-standing questions regarding the effectiveness of different ion transport mechanisms, providing guidance for rational design of fast ion-conducting media. In inorganic crystalline conductors, we find that ion transport occurs through both single-ion hops and aligned or unaligned concerted hops, depending on the temporal scale. Aligned concerted hops are up to twice as effective per event as single-ion hops. This concerted motion is particularly effective because it corresponds to macrostate transitions involving activated states, where lithium ions occupy high-energy positions due to the presence of nearby lithium neighbors. Therefore, in crystalline conductors, we reaffirm that raising the site energies of lithium ions through design strategies such as lithium stuffing can drastically improve ion transport.

Among polymer electrolytes, our analysis quantitatively shows that most Li transport in PEO arises from vehicular motion and complete or partial intrachain hops, both of which are limited by the sluggish Rouse dynamics of polymer segmental motion. Interestingly, although the polymer matrix in PPM exhibits slower segmental dynamics which might intuitively suggest lower ionic conductivity, Li ions in PPM can access interchain-hop events far more frequently. These highly effective pathways compensate for the slower polymer motion and yield overall higher Li diffusivity.

Our results highlight that tailoring monomer chemistry or designing copolymers with specific sequences may open new opportunities to further enhance ion transport by modulating the accessibility of these microscopic events. Moreover, our analysis suggests two design strategies for improving Li-ion conductivity in dry homopolymers. First, increasing the frequency of complete interchain hops could enhance ionic conductivity by more than fivefold. Second, although a less effective route, increasing the frequency of intrachain-hop events relative to vehicular events could also provide meaningful improvements.

Our results uncover fundamentally distinct mechanistic pathways governing ion transport in the two liquid electrolytes. In FAN, the overall diffusion is dominated by relatively slow AGG states, yet a small population of SSIP and CIP states undergoes highly effective solvent-exchange processes that disproportionately enhance Li-ion transport. In contrast, transport in EC arises primarily from a balance of slightly more effective solvent exchange and less effective vehicular motion of SSIP, with anion exchange contributing only marginally. This contrast illustrates how distinct solvation environments, and their associated transition pathways, shape the overall effectiveness of ion transport in liquid electrolytes. A design strategy to accelerate ion transport in liquid electrolytes is to weaken Li–solvent interactions just enough to enable rapid solvent exchange without reducing the degree of lithium salt dissociation, while also selecting low-viscosity solvents to enhance vehicular transport.

Furthermore, our analysis reveals an overarching picture of transport across all types of conductors, in which local environments mediate lithium migration pathways by stabilizing transition states. In polymer electrolytes, we find that interchain hops are more effective when accompanied by anion exchange, and in liquid electrolytes, anion exchange is more effective when mediated by surrounding solvents (when solvent exchange occurs simultaneously). When a lithium ion hops from one polymer chain to another, or exchanges an anion with its environment, the free-energy profile of the process is stabilized by the local solvent response. This concept is analogous to the ion-transport phenomenon known as the soft-cradle effect in inorganic plastic-crystalline conductors\citep{JunCederProceedingsoftheNationalAcademyofSciences2024}, where lithium-ion hops are stabilized by small-angle tilting of nearby polyanions. The recently proposed ligand-channel mechanism reflects the same principle, and a similar idea has been suggested for metal–organic framework conductors\citep{hou2022ionic-984}, where ion transport along the MOF wall must be mediated by surrounding solvent molecules.

Beyond what has been demonstrated in this work, the definitions of coordination microstates, macrostates and transitions can be systematically varied to enable a deeper insight. For example, the macrostate definition AGG can be further subdivided to number of charged particle in the aggregate to obtain deeper understanding in how transitions between aggregates of different sizes contribute to lithium transport. Furthermore, by tuning the distance cutoff in determining the coordination microstate, we can obtain mechanistic insights on transport as a function of spatial correlations.

Our framework validates long-standing microscopic models while revealing new mechanistic details in a fully quantitative manner. In inorganic crystalline conductors, event-resolved Arrhenius analysis confirms the lower activation barrier of concerted motion, long hypothesized from NEB studies. In polymers, sluggish Li mobility arises because transport is dominated by vehicular motion; interchain hops are highly effective but rare, while rolls (intrachain hops) are only marginally more effective than vehicular motion. In liquids, solvent and anion exchange operate on distinct timescales, explaining the exceptionally high ionic conductivity of FAN electrolytes. Together, these results consolidate prior qualitative hypotheses and, by enabling systematic comparison across chemistries, structures, and temperatures, provide a unified, quantitative picture of ion transport.

Beyond established systems, this framework opens a route to mechanism-aware electrolyte design. Because it requires only molecular dynamics trajectories and chemistry-informed definitions of microstates, macrostates, and transitions, it is readily transferable to emerging chemistries where mechanistic intuition is scarce. By isolating rate-limiting processes, quantifying their effectiveness, and pinpointing the temporal and spatial scales on which they operate, our computational framework can guide targeted modifications in chemistry and structure. Coupled with high-throughput computational workflows, it enables a closed design loop: diagnose transport bottlenecks, propose strategies to overcome them, perform large-scale simulations, analyze the resulting trajectories to extract further quantitative insight, and iterate. Although demonstrated here for Li conductors, the approach extends naturally to Na, protons, multivalent ions, and even mixed ionic–electronic systems, broadening its impact across electrochemical energy storage.

%%%%%%%%%%%%%%%%%%%%%%%%%%%%%%%%%%%%%%%%%%%%%%%%%%%%%%%%%%%%%%%%%%%%%%%%%%%%%%%%
% Methods
%%%%%%%%%%%%%%%%%%%%%%%%%%%%%%%%%%%%%%%%%%%%%%%%%%%%%%%%%%%%%%%%%%%%%%%%%%%%%%%%

\section{Methods}
\label{sec:methods}

\subsection{Classical Molecular Dynamics Simulations}
\paragraph{Force-field parametrization}
All polymer and liquid electrolyte simulations were performed using an OPLS-AA--based class-I force field\citep{jorgensen1988opls-dc4, jorgensen2005potential-2e9}. Force-field parameterization, system construction, and classical MD workflows were orchestrated using the HiTPoly repository\citep{wcn}, which interfaces with Packmol\citep{martnez2009packmol-a3a}, LigParGen\citep{dodda2017ligpargen-647}, and OpenMM\citep{eastman2017openmm-370}. Parameters for LiTFSI were taken from Leon \textit{et al.}\citep{LeonGomez-BombarelliChemistryofMaterials2025} (nitrogen-coordinating TFSI point charges), whereas \ce{LiPF6} and \ce{LiFSI} parameters were obtained from the CL\&P database by Padua \textit{et al.}\citep{lopes2004modeling-a74} Polymer repeat units (PEO, PPM) and solvents (EC, FAN) were parameterized using the LigParGen server. Partial atomic charges on polymers and solvents were computed using the Restrained Electrostatic Potential (RESP) scheme\citep{woods2000restrained-8a8}. To generate RESP charges, we constructed an ensemble of 50 conformations for each solvent or 3--6\,mer polymer fragment via short vacuum MD trajectories carried out in the NVT ensemble using initial OPLS-AA parameters obtained from LigParGen\citep{dodda2017ligpargen-647}. Single-point electronic-structure calculations were then performed in ORCA\citep{neese2012orca-de2} using the $\omega$B97X-D functional with the def2-TZVP basis set and D3 dispersion corrections, and RESP charges were extracted using Multiwfn\citep{lu2012multiwfn-f89}.

Because class-I force fields lack explicit polarizability, all atomic charges on Li and TFSI in polymer electrolytes were uniformly scaled by 0.7, following prior work on polymer electrolytes\citep{molinari2018effect-7bb, france-lanord2020effect-522, ruza2025benchmarking-d60, LeonGomez-BombarelliChemistryofMaterials2025}. This scaling improves agreement with experimental conductivities at moderate-to-high salt loadings and prevents the over-binding and ion clustering observed with unscaled or semiempirical LigParGen charges. For liquid electrolytes, Li and \ce{PF6}/\ce{FSI} atomic charges were uniformly scaled by 0.8, following the approach of Hou \textit{et al.}\citep{hou2021solvation-ec1}

\paragraph{System construction and simulations}
Polymer electrolyte systems consisted of 30 chains, each containing 100 and 30 repeat units for PEO and PPM, respectively and terminated with methyl groups. For polymer systems, Li and TFSI ions were added to achieve Li:EO ratios of $r = 0.01,\ 0.033,\ 0.06,\ 0.08,\ 0.10,$ and $0.12$ (with $r = 0.08$ used only for PPM), while maintaining charge neutrality. In the liquid electrolyte systems, the EC electrolyte contained 113 Li–\ce{PF6} ion pairs and 1500 EC molecules (1.0 M), whereas the FAN electrolyte contained 141 Li–FSI ion pairs and 1700 FAN molecules (1.3 M). All initial configurations were generated with Packmol and equilibrated using the multistep protocol described by Ruza \textit{et al.}\citep{ruza2025benchmarking-d60} All MD simulations were carried out using OpenMM with GPU acceleration.

Following equilibration, production simulations of 500\,ns for polymer electrolytes and 100\,ns for liquid electrolytes were performed. All simulations used a 12\,\AA\ cutoff for van der Waals interactions and short-range electrostatics, with long-range Coulomb interactions treated using the particle-mesh Ewald method. Because classical MD systematically overestimates polymer glass transition temperatures, we evaluated the $T_g$ of PEO and PPM at $r = 0.08$ and found offsets of 20 K and 90 K, respectively (see Supplementary Information \ref{si:tg}). Accordingly, following Fang \textit{et al.}\citep{fang2023molecular-43c}, simulations for PEO and PPM were performed at shifted temperatures of 373 K and 443 K to effectively reproduce experimental behavior at 80$^\circ\mathrm{C}$ (353 K). Liquid electrolyte production simulations were performed in the NVT ensemble at 298 K. Temperature was controlled using a Nos\'e--Hoover thermostat with a chain length of 3, and all hydrogen-containing bond lengths were constrained. A timestep of 2\,fs was used for all simulations. Atomic coordinates were saved every 2\,ps for liquid electrolytes and every 10\,ps for polymer electrolytes.

\subsection{Fine-tuning machine-learning interatomic potentials}
\paragraph{MLIP training}
We used atomic configurations of \ce{Li_{1+x}Al_xTi_{2-x}(PO4)3} (LATP) at $x = 1/3$ from MD simulation frames taken from \citet{jun2025exploring-f0b}, simulated using the Perdew--Burke--Ernzerhof (PBE) generalized gradient approximation (GGA) exchange-correlation functional \citep{PerdewErnzerhofPhysicalReviewLetters1996} at 300 K and 600--1200 K. From all raw MD snapshots, we used a method similar to the DIRECT sampling approach \citep{qi2024robust-d1d,kaplan2025foundational-226} to downsample the dataset and ensure structural diversity between the train and test split. We extracted invariant embeddings from two layers of the pretrained MACE-MPA-0 model (mace-torch 0.3.10 package) \citep{batatia2023foundation-d3c} and normalized them across the dataset. Principal Component Analysis (PCA) was applied to select 16 features with explained variance greater than 1 (Kaiser rule \citep{kaiser1960application-ac8}), and the selected features were scaled by their explained variance ratio. BIRCH clustering \citep{zhang1996birch-1a1} with a threshold of 0.15 was then applied, resulting in 34,278 structural clusters. Dimensionality reduction and clustering were performed with the scikit-learn 1.6.1 package \cite{PedregosaDuchesnayarXiv2012}. The cluster representatives were randomly selected and randomly divided into 90/10 train/test sets (30,851/3,428 structures). Then, we fine-tuned the MACE-MPA-0 model on this dataset over 10 epochs with a batch size of 8, a learning rate of 0.0001, and energy and force weights both set to 1.0. Among the training data, 5\% was used as a validation split, and the multi-head fine-tuning option was disabled. The fine-tuned MACE-MPA-0 model achieved a test set MAE of 0.14 meV/atom (energy) and 14.4 meV/\AA{} (forces).

\paragraph{MD simulation of inorganic crystalline materials}
For three lithium concentrations of LATP at $x = 0$, $1/6$, and $1/3$, the lithium ion ordering was initialized following \citep{XiaoCederAdvancedEnergyMaterials2021}. The simulation boxes were scaled up to contain 432, 436, and 440 atoms, with 24, 28, and 32 lithium atoms, respectively, for each lithium concentration. For each temperature in 800, 850, 900, $\cdots$, 1100 K, initial velocities were drawn from Maxwell--Boltzmann distributions. Each system then underwent 20 ps (10k steps) equilibration, followed by 10 ns (5M steps) production runs under the NVT ensemble at the target temperature using a time step of 2 fs. The temperature was controlled using the Nos\'e--Hoover chains thermostat \citep{martyna1992noshoover-853} as implemented in the ASE 3.25.0 package \citep{larsen2017atomic-f97}, with thermostat time constants of 100 fs for equilibration and 200 fs for production. Atomic trajectories were saved every 0.1 ps (50 steps).

For NASICON materials, temperatures higher than the ones included in our studies encountered a permutational 120 degree rotational event of phosphate group, as reported in an earlier work \cite{jun2025exploring-f0b}. Our assignment of coordination site uses a static list of coordinating oxygens for each 6b, 18e, 36f sites in Li-NASICONs, and these rare large-angle rotation events results in a permutation of oxygen indices. While modification of the index of coordinating oxygens can be performed whenever we detect a 120 degree rotation event, for simplicity in our work, we adhere to low-temperature simulations where such permutative rotation events do not occur at all during our simulations.

%\section*{Data Availability}

\section*{Code Availability}
The code to reproduce this work is available on GitHub: \url{https://github.com/learningmatter-mit/OnsagerDecomposer}.
%Change the name of the package to OnsagerDecomposer and make it public.

\section*{Author Contributions}
K.J. conceptualized the project and developed the theoretical framework with P.L. and R.G.-B. J.R. and P.L. parametrized forcefields for liquid and polymer electrolytes. K.J. performed MD simulations for liquids and polymer electrolytes. J.N. fine-tuned MLIP and performed MD simulations of crystalline conductors. All authors reviewed and edited the paper.

\begin{acknowledgments}
This work was supported by the Energy Storage Research Alliance "ESRA" (DE-AC02-06CH11357), an Energy Innovation Hub funded by the U.S. Department of Energy, Office of Science, Basic Energy Sciences. We acknowledge the MIT Lincoln Laboratory Supercloud clusters as well as computational resources of the National Energy Research Scientific Computing Center (NERSC). The authors acknowledge partial support from the Toyota Research Institute.
\end{acknowledgments}

\bibliographystyle{apsrev4-2}
\bibliography{main_clean}

%%%%%%%%%%%%%%%%%%%%%%%%%%%%%%%%%%%%%%%%%%%%%%%%%%%%%%%%%%%%%%%%%%%%%%%%%%%%%%%%
% Extended Data
%%%%%%%%%%%%%%%%%%%%%%%%%%%%%%%%%%%%%%%%%%%%%%%%%%%%%%%%%%%%%%%%%%%%%%%%%%%%%%%%

\widetext
\clearpage

\setcounter{figure}{0}
\setcounter{table}{0}
\renewcommand{\figurename}{Extended Data Fig.}
\renewcommand{\tablename}{Extended Data Table}
\crefalias{figure}{extendedfigure}
\crefalias{table}{extendedtable}

\label{fig:corrector_ablation}

% \widetext
\clearpage

\begin{center}
\textbf{\large Supplementary Information for: \\[1ex] Universal Framework for Decomposing Ionic Transport into Interpretable Mechanisms}
\end{center}

\setcounter{equation}{0}
\setcounter{figure}{0}
\setcounter{table}{0}
\setcounter{section}{0}
\setcounter{page}{1}
\renewcommand{\theequation}{S\arabic{equation}}
\renewcommand{\thefigure}{S\arabic{figure}}
\renewcommand{\thetable}{S\arabic{table}}
\renewcommand{\thesection}{\Alph{section}}
\renewcommand{\thesubsection}{\arabic{subsection}}
\renewcommand{\figurename}{Fig.}
\renewcommand{\tablename}{Table}
\crefalias{figure}{suppfigure}
\crefalias{table}{supptable}
\crefalias{section}{suppsection}

{
\section*{Table of Contents}
\setstretch{2.0}
\contentsmargin{2.55em}
\dottedcontents{section}[3.8em]{}{2.3em}{1pc}
\dottedcontents{subsection}[7.6em]{}{2.3em}{1pc}

\startcontents
\printcontents{}{1}{}{}
}
\clearpage
\section{Efficient computation of full time-window-averaged correlations}
\label{si:fft_computation}

Direct computation of time-window-averaged correlations between particles $i$ and $j$ as a function of the window length scales poorly with the total number of simulation steps $N$, exhibiting $O(N^2)$ computational cost. In particular, the \textit{OnsagerDecomposer} algorithm decomposes the motion of each particle into $n$ transport modes by representing them as virtual particles, which makes full time-window averaging of displacements computationally intractable. For a window size corresponding to $\Delta t = n_i$ steps, $N - n_i + 1$ windows must be evaluated and averaged over the full trajectory. In practice, this computational burden is often reduced by averaging only a subset of windows or by evaluating correlations at a limited set of $\Delta t$ values. However, such approximations underutilize trajectory data and may lead to noisier displacement correlation estimates.

To address this limitation, we employ a Fast Fourier Transform (FFT)-based algorithm to perform full time-window averaging efficiently\citep{calandrini2011nmoldyn-446}. This algorithm is implemented in the \textit{OnsagerDecomposer} package used throughout this work. This approach enables the use of the entire trajectory while reducing the computational scaling to $O(N \log N)$, allowing correlations to be computed reliably across a wide range of $\Delta t$ values. The correlation between particles $i$ and $j$ in Eq.~S1 can be decomposed into two contributions, denoted $S_1$ and $S_2$. The $S_1$ term can be evaluated with linear scaling, $O(N)$, while $S_2$ is computed using FFT-based convolution, resulting in overall $O(N \log N)$ complexity. The implementation has been verified to reproduce results obtained from direct summation over $\Delta t$ and $k$ for both self and distinct particle correlations.

\begin{equation}
    \text{Corr}_{i,j}(\Delta t) =
    \frac{1}{N-\Delta t+1}\sum_{k=0}^{N-\Delta t}
    \left( r_i(k+\Delta t) - r_i(k) \right)
    \cdot
    \left( r_j(k+\Delta t) - r_j(k) \right)
    = \frac{1}{N-\Delta t+1} \left(S_1 + S_2\right)
\end{equation}

\begin{equation}
    S_1 =
    \sum_{k=0}^{N-\Delta t}
    \left[
        r_i(k+\Delta t) \cdot r_j(k+\Delta t)
        +
        r_i(k) \cdot r_j(k)
    \right]
\end{equation}

\begin{equation}
    S_2 =
    -\sum_{k=0}^{N-\Delta t}
    \left[
        r_i(k) \cdot r_j(k+\Delta t)
        +
        r_j(k) \cdot r_i(k+\Delta t)
    \right]
\end{equation}

\clearpage
\section{Decomposition of self and distinct displacement correlations}
\label{si:self_distinct_correlations}

The \textit{OnsagerDecomposer} algorithm can decompose Onsager transport coefficients into contributions associated with user-defined microstate or macrostate transitions. However, as reported in prior literature, achieving statistical convergence of distinct particle correlations within feasible simulation times remains challenging. Moreover, since the algorithm further decomposes self and distinct correlations into contributions from individual transitions, convergence becomes limited by the rarest transition events, making convergence of decomposed correlations even more difficult than that of conventional distinct particle correlations.

\begin{figure*}[!ht]
\centering
\includegraphics[width=1.0\textwidth]{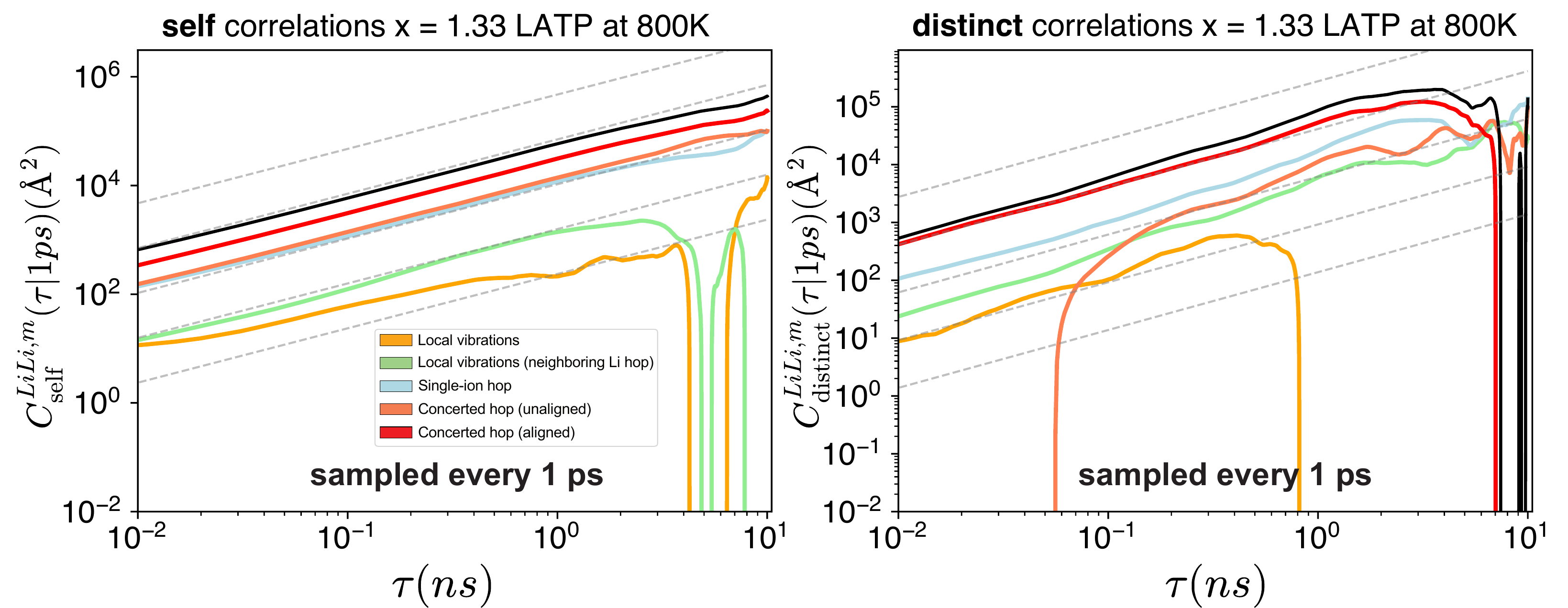}
\caption{
\textbf{Decomposition of self and distinct displacement correlations in $x = 1.33$ LATP at 800~K.}
The left panel shows decomposition of self Li--Li correlations, while the right panel shows decomposition of distinct Li--Li correlations in $x = 1.33$ \ce{Li_{1+x}Al_{x}Ti_{2-x}(PO4)3} at 800~K using a sampling interval of 1~ps. Black lines denote total self or distinct correlations, and colored lines represent the corresponding decomposed contributions.
}
\label{fig:si_correlations}
\end{figure*}

As an example, Fig.~\ref{fig:si_correlations} shows the decomposition of self and distinct displacement correlations in Li-stuffed LATP. The total self correlation (black line in the left panel) and its decomposed components (colored lines) exhibit linear growth with respect to $\tau$, indicating good convergence. In contrast, the distinct Li--Li correlation (black line in the right panel) is noticeably noisier, and the decomposed distinct correlations do not achieve satisfactory convergence within the 10~ns simulation time. Consequently, the present study primarily reports decompositions of self Li--Li correlations. Nevertheless, for trajectories that are an order of magnitude longer, the \textit{OnsagerDecomposer} framework can reliably provide mechanistic decomposition of distinct particle correlations, which we leave for future investigation.

\clearpage
\section{OnsagerDecomposer results for stoichiometric \ce{LiTi2(PO4)3}}
\label{si:stoichiometric_ltp}

\begin{figure*}[!ht]
\centering
\includegraphics[width=1.0\textwidth]{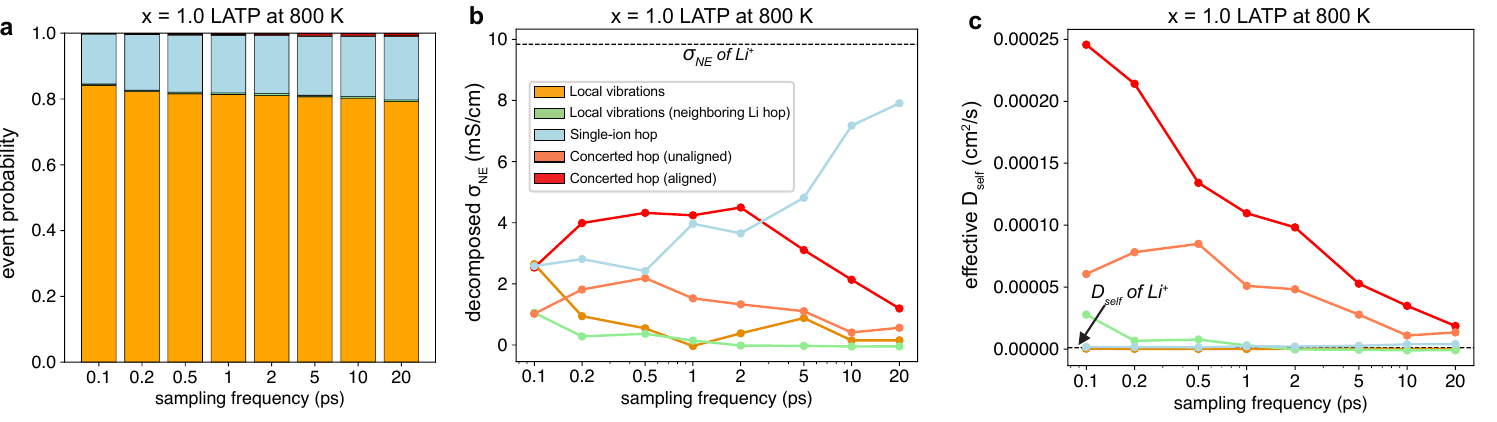}
\caption{
\textbf{Mechanistic decomposition of transport in stoichiometric \ce{LiTi2(PO4)3} at 800~K.}
\textbf{a}, Probability of events \(p_m(\Delta t)\) versus sampling window \(\Delta t\) at 800 K.
\textbf{b}, Contributions of the various microstate transition types to \(D^{\mathrm{Li}}_{\mathrm{self}}\) as a function of \(\Delta t\). The sum over events at each \(\Delta t\) equals the undecomposed \(D^{\mathrm{Li}}_{\mathrm{self}}\).
\textbf{c}, Effective self-diffusion for each event, defined as \(D^{\mathrm{Li}}_{\mathrm{self},m}/p_m(\Delta t)\), as a function of \(\Delta t\).
}
\label{fig:si_stoichiometric_ltp}
\end{figure*}

\clearpage
\section{Conventional transport analysis for PEO and PPM polymer electrolytes}
\label{si:correlation_polymer}

\begin{figure*}[!ht]
\centering
\includegraphics[width=1.0\textwidth]{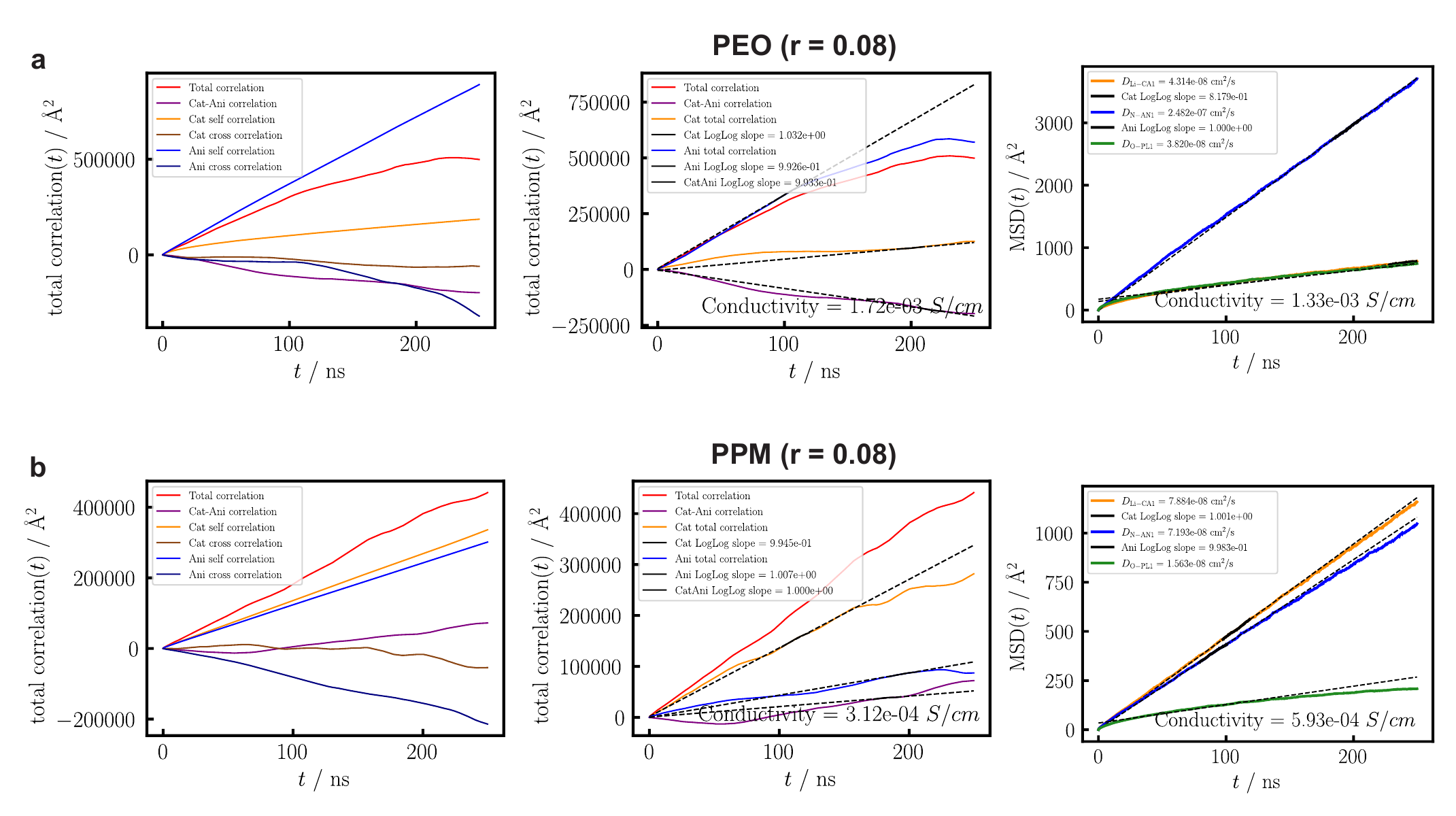}
\caption{
\textbf{Conventional transport coefficients for polymer electrolytes.}
Green--Kubo correlation functions used to compute Onsager transport coefficients and mean-squared displacements for PEO ($r = 0.08$) (a) and PPM ($r = 0.08$) (b). These quantities were computed using analysis tools provided in the HiTPoly repository. $\mathrm{D_{O-PL1}}$ in the right panels corresponds to the self-diffusion coefficient of oxygen atoms in polymer chains, which we use as a proxy for constitutional repeat unit (CRU) diffusivity.
}
\label{fig:correlations_polymer}
\end{figure*}

Figure \ref{fig:correlations_polymer} presents conventional Onsager transport coefficients for PEO and PPM ($r = 0.08$) without decomposition into mechanistic components. All quantities are computed using analysis tools provided in the HiTPoly repository\citep{ruza2025benchmarking-d60}. ``Cat'' and ``Ani'' denote cations (Li) and anions (TFSI), respectively. The left panels report Li--Li self and distinct correlations, anion--anion self and distinct correlations, and Li--anion distinct correlations. These contributions are combined to yield total cation and anion correlations shown in the center panels. The right panels display mean-squared displacements of Li (denoted Li-CA1), anions (denoted N-AN1), and polymer constitutional repeat units (CRUs, denoted O-PL1), obtained by tracking Li, N, and O atoms, respectively.

These correlations highlight fundamental differences between PEO and PPM electrolytes. Comparison of $\mathrm{D_{CRU}}$ shows that PEO chains move more than twice as fast as PPM chains. In PEO, Li self-diffusivity $\mathrm{D_{self}}$ closely follows $\mathrm{D_{CRU}}$, whereas in PPM, Li $\mathrm{D_{self}}$ is significantly larger than $\mathrm{D_{CRU}}$. This observation is consistent with the mechanistic analysis obtained using OnsagerDecomposer: Li transport in PEO is dominated by vehicular motion and intrachain hopping with minimal interchain hopping, while Li transport in PPM primarily occurs through interchain hopping, resulting in strong decoupling between Li motion and polymer dynamics.

\clearpage
\section{Cluster population in EC and FAN-based liquid electrolytes}
\label{si:cluster_population_liquids}

\begin{figure*}[!ht]
\centering
\includegraphics[width=1.0\textwidth]{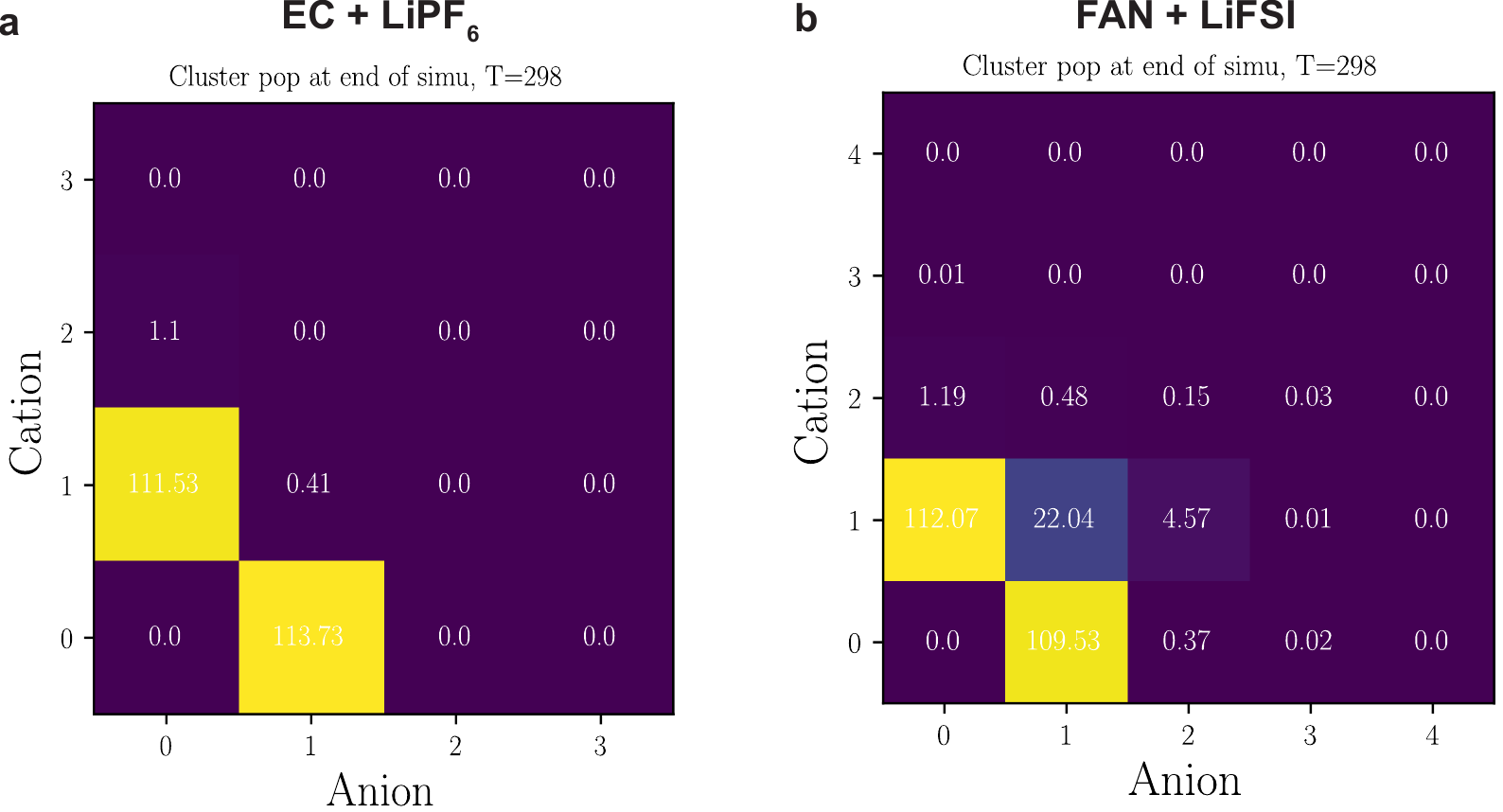}
\caption{
\textbf{Cluster population map}.
Cluster population matrices at the end of the simulation for EC with \ce{LiPF6} and FAN with LiFSI. Each matrix element denotes the number of clusters containing $i$ cations and $j$ anions. While the EC electrolyte contains negligible populations of contact ion pairs and aggregates, the FAN electrolyte exhibits a substantial fraction of contact ion pairs and aggregates.
}
\label{fig:cluster_population}
\end{figure*}

\clearpage
\section{Determining glass transition temperature offsets for polymer electrolytes}
\label{si:tg}

Classical MD simulations are known to systematically overestimate polymer glass transition temperatures. Therefore, when quantitative comparison of ionic conductivities with experiments is required, simulation temperatures are conventionally shifted to account for the offset between computed and experimental $T_g$ values.

The glass transition temperature is obtained from density measurements performed in the NPT ensemble during a temperature sweep from 500 K to 100 K in increments of 20 K. At each temperature, the system is equilibrated for 10 ns, and the density is averaged over the final 2 ns. The $T_g$ is determined from the intersection of bilinear fits to the high-temperature rubbery regime and low-temperature glassy regime. This protocol is consistent with prior studies\citep{ruza2025benchmarking-d60, fang2023molecular-43c}.

Experimentally measured $T_g$ values for PEO and PPM at $r = 0.08$ are 233 K and 250 K, respectively\citep{fang2023molecular-43c}. As shown in Figure Our simulations yield $T_g$ values of 253 K for PEO and 339 K for PPM, corresponding to offsets of 20 K and 90 K, respectively. Accordingly, to model transport at 80~$^\circ\mathrm{C}$ (353~K), simulations are performed at 373 K for PEO and 443 K for PPM.

\begin{figure*}[!ht]
\centering
\includegraphics[width=1.0\textwidth]{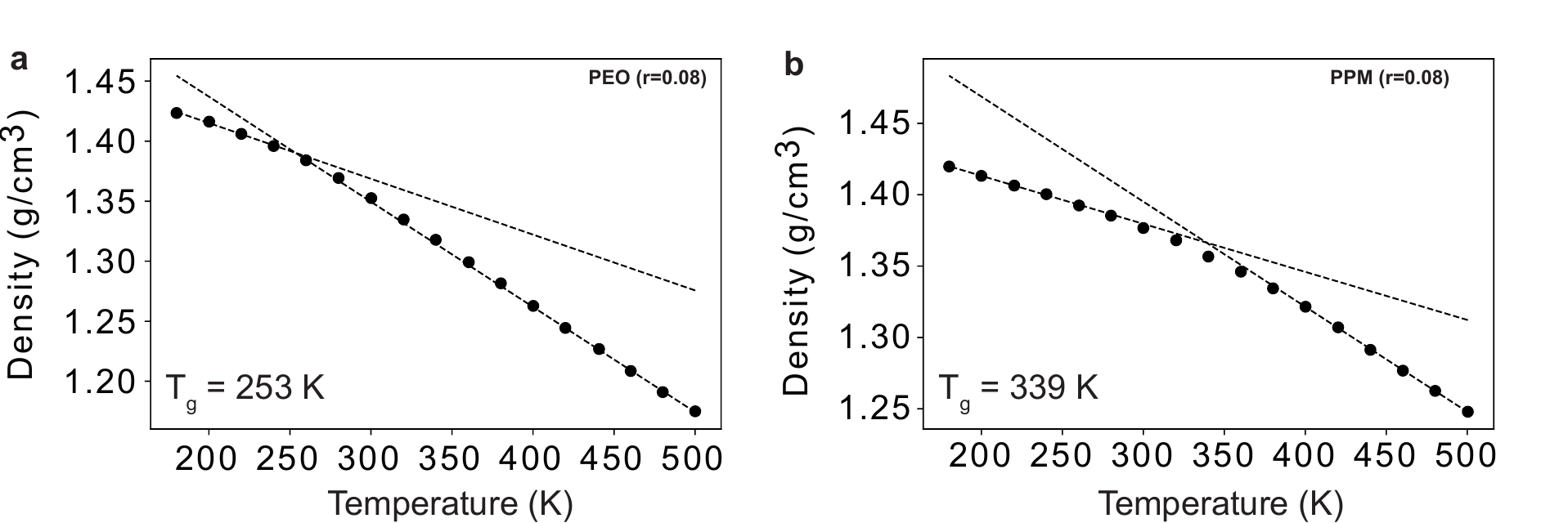}
\caption{
\textbf{Computing glass transition temperatures}.
Computed glass transition temperatures of PEO (a) and PPM (b) at r = 0.08. Bilinear fits yield $T_g$ values of 253 K and 339 K for PEO and PPM, respectively; these values are used to shift simulation temperatures to match experimental $T_g$.
}
\label{fig:tg}
\end{figure*}

\end{document}